\documentclass[preprint]{elsarticle}
\usepackage{aas_macros}
\usepackage[a4paper,centering, totalwidth=520pt, totalheight=700pt]{geometry}

\def\k#1 {k_{{\rm #1}}}

\def\mH2p{H_2^+}

\def\ltsima{$\; \buildrel < \over \sim \;$}
\def\simlt{\lower.5ex\hbox{\ltsima}}   
\def\gtsima{$\; \buildrel > \over \sim \;$}
\def\gtsim{\lower.5ex\hbox{\gtsima}}

\usepackage{amsmath} 
\usepackage{graphicx}
\usepackage{amssymb}
\usepackage{amsmath}
\usepackage{enumerate}
\usepackage{microtype}
\usepackage{color}
\usepackage[caption=false,listofformat=subparens]{subfig}
\usepackage{mathtools}

\renewcommand{\v}[1]{\ensuremath{\boldsymbol{\mathbf{#1}}}} 

\newcommand{\pd}[2]{\frac{\partial #1}{\partial #2}}

\let\baraccent=\= 
\renewcommand{\=}[1]{\stackrel{#1}{=}} 

\newcommand{\comment}[1]{}

\newcommand{\p}{\partial}
\newcommand{\vp}{Vlasov--Poisson }

\begin{document}
\begin{abstract}

We extend the simplex-in-cell (SIC) technique recently introduced in the context
of collisionless dark matter fluids~\citep{Abel:2012,Hahn:2012} to the case of
collisionless plasmas. The six-dimensional phase space distribution function
$f(\mathbf x,\mathbf v)$ is represented by an ensemble of three-dimensional
manifolds, which we refer to as sheets. The electric potential field is obtained
by solving the Poisson equation on a uniform mesh, where the charge density is
evaluated by a spatial projection of the phase space sheets. The SIC
representation of phase space density facilitates robust, high accuracy
numerical evolution of the Vlasov-Poisson system using significantly fewer
tracer particles than comparable particle-in-cell (PIC) approaches by reducing
the numerical shot-noise associated with the latter. We introduce the SIC
formulation and describe its implementation in a new code, which we validate
using standard test problems including plasma oscillations, Landau damping, and
two stream instabilities in one dimension. Merits of the new scheme are shown to
include higher accuracy and faster convergence rates in the number of
particles. We finally motivate and outline the
efficient application of SIC to higher dimensional problems.

\end{abstract}

\title{Simplex-in-Cell Technique for Collisionless
  Plasma Simulations}

\author[slac]{Julian Kates-Harbeck\fnref{curraddress}}
\author[slac]{Samuel Totorica}
\author[slac]{Jonathan Zrake}
\author[slac,lagrange]{Tom Abel}
\ead{tabel@stanford.edu}

\cortext[cor]{Corresponding Author}
\fntext[curraddress]{Present address: Department of Physics, Harvard University, Cambridge, MA 02138, U.S.A.}
\address[slac]{Kavli Institute for Particle Astrophysics and Cosmology,  
  Stanford University, \\ \quad SLAC National Accelerator Laboratory,
  Menlo Park, CA 94025, USA}
\address[lagrange]{Institut Lagrange de Paris, Institut d'Astrophysique de Paris,
  98 bis boulevard Arago - 75014 Paris, France}

%

\maketitle


\section{Introduction}
Many astrophysical and laboratory plasmas of interest are characterized by
densities and temperatures in the collisionless regime. In collisionless plasmas
the collective effects resulting from the long-ranged electromagnetic fields
dominate the effects of two-body collisions, to the extent that such local
interactions may be neglected \cite{Chen:1984}. The foundational fully kinetic
description of a collisionless plasma is given by the Vlasov equation, which
when coupled with Maxwell's equations describes a continuum of charged particles
that self-consistently evolve under the electromagnetic fields they
generate. When particles and plasma waves all move slowly compared to light,
couplings are effectively electrostatic and the complete system is referred to
as Vlasov-Poisson. Obtaining solutions to this system is essential for the
understanding of many of the problems of modern interest in plasma physics, such
as astrophysical particle acceleration \cite{Cerutti:2014, Sironi:2014},
turbulence and heating in the solar wind and corona \cite{Pueschel:2014,
  Valentini:2014} and laser-plasma interactions \cite{esarey2009physics,Fiuza:2013,
  Kemp:2014}. The nonlinear nature of the Vlasov-Poisson system makes it
intractable to solve analytically in non-trivial scenarios; obtaining results
under realistic conditions is thus generally approached using computer
simulations.

The development of methods for solving the Vlasov-Poisson system numerically has
been an active area of research for over 50 years \cite{Dawson:1962}. Although a
great diversity of solution schemes exist, they all adhere to the same basic
blueprint, which involves an approximate representation of the phase-space
density $f(\mathbf x, \mathbf v)$ of the charge carriers, determining the charge
density $\rho$ from that representation, recovering the electric field from the
elliptic constraint $\nabla \cdot \mathbf E = 4\pi \rho$, and finally using
$\mathbf E$ to advance $f(\mathbf x, \mathbf v)$ in time. The various solution
schemes differ most dramatically in their representation of the phase-space
density. These fall into two basic categories \cite{Denavit:1971}: the
particle-based (or ``Lagrangian'') methods and the grid-based Eulerian methods
(so-called ``Vlasov codes''). The particle-based methods represent
$f(\mathbf{x}, \mathbf v)$ as the superposition of compactly supported
probability mass elements (macro-particles) that move about the phase-space
along characteristics of the Vlasov equation. Meanwhile, the grid-based Eulerian
codes tabulate $f(\mathbf x, \mathbf v)$ on a multi-dimensional mesh extending
over relevant portions of the phase-space. Both approaches are explicitly
conservative with respect to total probability mass, but otherwise each offers
distinct merits and limitations that must be weighed carefully when choosing the
optimal computational approach to a given problem. The scheme which we introduce
in this paper adopts a novel representation of $f(\mathbf x, \mathbf v)$ that
accommodates favorable aspects of both approaches and can be seen as a hybrid;
phase-space is covered (partially) by an ensemble of three-dimensional meshes,
whose zero-dimensional \emph{vertices} evolve along the Vlasov characteristics
ensuring that probability mass is invariant within each finite volume. Before
describing our new formulation in detail, we will briefly review relevant
aspects of the particle and grid-based approaches.

The particle-in-cell approach (PIC), though very successful in yielding insight
into the complex nonlinear physics governing collisionless plasmas
\cite{birdsall2004plasma, hockney1988computer, Dawson:1983}, does have its
limitations. For example, the macro-particles can suffer from artificial
pairwise interactions, which are not present in the collisionless Vlasov-Poisson
system \cite{Okuda:1970}. This artificial collisionality can be reduced, though
not eliminated, by an appropriate choice for the particle shape and size
\cite{Langdon:1970}. Further limitations of the PIC methods include their
failure to capture solutions features whose amplitudes are exceeded by the level
of Poisson noise. This danger can only be avoided by increasing the number of
macro-particles, which comes at a large computational cost --- a value of the
distribution function described by $N$ particles in a volume of phase-space has
an error proportional to $N^{-1/2}$ \cite{Hammersley:1965}. Thus, achieving an
acceptable signal-to-noise ratio in PIC simulations often requires prohibitively
large computational resources. Limited statistical sampling is, by construction,
most severe in regions of low phase-space density (e.g. tails of the velocity
distribution) which are crucial for resolving kinetic effects, such as
wave-particle interactions. An alternative method of initializing the particles
that avoids statistical under-sampling and the associated noise is the ``quiet
start'' method \cite{Byers:1970}. Using a quiet start, the velocity distribution
is initialized with particle velocities that are equally separated in velocity
space. The particles of a given species will still have identical charge to mass
ratios, but the charges of individual particles now vary according to the
velocity distribution. (Alternatively, one may use particles with identical
charges but separated unequally in velocity space, with the intervals chosen to
correctly approximate the velocity distribution.) The quiet start method is free
from statistical fluctuations in the initial state of the simulation, allowing
one to study small amplitude effects that may be washed out by noise when a
random initialization is used. However, quiet start initializations cannot
alleviate undersampling for non-linear processes, after the system has evolved
far away from its initial state. The quiet start is also subject to numerical
instabilities associated with the interaction between the different ``streams''
in velocity space \cite{dawson1960plasma}.

Eulerian methods directly solve the Vlasov equation on a discretized grid in
phase-space.  This has the advantage that the distribution function is equally
well resolved in all regions of phase-space, allowing the simulation to capture
wave-particle interactions much easier than with PIC.  Eulerian methods are also
free from the random fluctuations resulting from the use of discrete particles.
This gives them a very low noise level, enabling the resolution of fine details
in phase-space that may be obscured by the noise inherent to PIC.  Many true
solutions to the Vlasov equation exhibit ``velocity filamentation''
\cite{Knorr:1973}, where structure in velocity space is produced on increasingly
fine scales as time progresses.  Eulerian methods are more easily able to capture such details than PIC, due to the better resolution in velocity space and the fact that they are not subject to an exponential sampling falloff for high particle speeds.
However, unless adaptive methods are employed, the filamentation will eventually
reach scales below the resolution of the Eulerian grid, and this information
will be lost.

Describing the distribution function on a discretized Eulerian grid however is
generally also very expensive computationally due to the high dimensionality of
phase-space (three spatial dimensions and three velocity dimensions.)  This
severely limits the spatio-temporal resolution feasible in Eulerian simulations.
This is one of the main drawbacks of the Eulerian method, and has restricted its
use mostly to problems that can be adequately described with reduced
dimensionality in phase-space.  A large variety of algorithms have been
developed in the effort to improve the efficiency of the Eulerian method while
retaining its reduced levels of noise.  These include standard numerical
integration techniques such as finite difference methods \cite{Whitson:1978},
finite element methods \cite{Zaki:1988}, and finite-volume methods
\cite{Sircombe:2009}.  Semi-Lagrangian methods \cite{Cheng:1976,Besse:2003},
where the distribution function is advected in a Lagrangian manner and
interpolated to an Eulerian phase-space grid, can free the Eulerian method from
the Courant-Friedichs-Lewy condition on the timestep.  Methods implementing
adaptively refined phase-space grids are able to resolve the fine detail in the
distribution function at all times throughout the simulation and can thus better
handle problems where resolution of velocity filamentation is necessary
\cite{Besse:2008}.  The grids used for Eulerian methods are limited not only to
the physical space and velocity dimensions; spectral methods have been developed
which use a Fourier decomposition of the spatial part of the distribution
function, and a Fourier \cite{Engelmann:1963} or Hermite
\cite{Schumer:1998,Camporeale:2013} decomposition of velocity space.
Fourier-Hermite decompositions of the distribution function are especially well
suited for the modeling of warm plasmas with near Maxwellian velocity
distributions, as a Maxwellian is represented exactly by the lowest order
Hermite polynomial.

With both particle-based and grid-based methods having their strengths and
weaknesses, the choice of which is best to use will depend on the details of the
problem in question such as the level of noise that is acceptable, and the
dimensionality and resolution required in velocity space \cite{Besse:2008}.
Despite its limitations, PIC is currently the best tool available for many
problems and remains the standard technique for simulating collisionless
plasmas.  The algorithms at the heart of the PIC method have remained unchanged
for decades, but there has been a large effort to develop improvements to PIC
that may reduce the numerical noise and make simulations more economical
\cite{Verboncoeur:2005}.  Promising methods to reduce the discrete particle
noise have been demonstrated, such as re-mapping algorithms
\cite{Denavit:1972,Wang:2011b} which periodically reconstruct the distribution
function throughout the simulation, and use this to create a new particle
discretization with improved noise properties.  When the distribution function
can be decomposed into a perturbation about a known analytical equilibrium
solution, $\delta f$ methods \cite{kotschenreuther1988deltaf,Dimits:1993} can reduce particle noise by
discretizing only the perturbed part of the distribution function. Algorithms that allow the use of various particle smoothing functions for the reduction of sampling noise while explicitly preserving symplectic structure \cite{xiao2013variational} or momentum and energy conservation \cite{evstatiev2013variational} have also been developed. The
stability limits on the time step and spatial resolution arising from the
explicit time integration scheme standard to PIC ($\omega_{p} \Delta t < 2$ and
$\Delta x < \lambda_{D}$) may be overcome through the use of implicit time
integration schemes \cite{markidis2011energy,Mason:1981,Chen:2011}, allowing the use of coarser
discretization in space and time and thus reducing the computational cost of the
simulation.  Hybrid methods that treat the ion component as particles but use a
fluid description of the electron component allow simulation on the longer ion
timescales for problems where the details of the high-frequency electron
dynamics are not important \cite{Mason:1971,Cartwright:2000}.  Beyond
algorithmic improvements, optimizing PIC has been approached from many other
angles including more efficient software implementations
\cite{Fonseca:2008,Fonseca:2013}, and tuning of applications for specific
hardware architectures \cite{Bowers:2009}.  These are considerations of
increasing importance for the heterogeneous systems emerging with the advent of
the exascale computational era.

The PIC method for plasma simulation is closely related to the $N$-body methods
used for cosmological simulations \cite{Efstathiou:1981, Springel:2005,
  Angulo:2012a}, which use a particle discretization to model the evolution of a
self gravitating, collisionless fluid of dark matter.  $N$-body methods commonly
employ a particle-mesh technique to compute the long-range gravitational forces
between particles, augmented with a direct summation or tree method to calculate
the short-range forces.  This helps to capture the large dynamic range necessary
in modeling cosmological scenarios.  Cosmic structure formation simulations
begin with a dark matter distribution that is nearly uniform in space, with
perturbations sampled from a power spectrum corresponding to the initial
conditions indicated by cosmic microwave background radiation measurements.  The
system then evolves via the gravitational instability to form the features that
make up the large-scale structure of the universe - the ``cosmic web'' - such as
filaments, halos, and voids.  The initial velocity dispersion of the dark matter
fluid is small compared to the bulk flow velocities that arise during its
subsequent evolution, and thus the fluid may be modeled as perfectly cold to a
good approximation.  The corresponding phase-space distribution function is then
initially single-valued in velocity at each point in position space, and thus
occupies a three-dimensional submanifold (or ``sheet'') within six-dimensional
phase-space.  As the system evolves, the phase-space sheet is distorted in a
complex manner without tearing, and the connectivity of points on the sheet does
not change.

\cite{Abel:2012} and \cite{Shandarin:2012} had the critical insight that the
simulation particles may be thought of as the vertices of an unstructured mesh
that traces the evolution of this phase-space sheet.  This allows the
construction of a piecewise approximation to the distribution function at any
time in the simulation by interpolating between these ``tracer particles''.  One
can then define density, bulk velocity, and velocity dispersion fields
continuously over the spatial domain and not subject to the noise and reduced
resolution inherent to averaging over control volumes.  As structures collapse
gravitationally during the simulation, the velocity field becomes multi-valued
where multiple ``streams'' of the collisionless dark matter fluid overlap in
position space.  The new method is able to resolve the individual contributions
of these streams, giving access to much more accurate and detailed estimates of
velocity fields and their differentials \cite{Hahn:2014}.  A new field that
counts the number of fluid streams at each location in space can also now be
defined, which has been demonstrated to yield new insight into cosmic structure
\cite{Abel:2012,falck2012origami}.

The phase-space sheet method was originally applied as a postprocessing tool to
reveal fine scale detail in simulations performed with standard N-body methods.
\cite{Kaehler:2013} demonstrated that the method may be used to create
visualizations that are free from the unphysical granularity seen when using
standard particle rendering techniques (such as adaptive kernel smoothing.)  The
phase-space sheet was implemented in the particle-mesh method to calculate the
density for Poisson's equation in \cite{Hahn:2013}.  They were able to
demonstrate promising improvements to the particle-mesh method, such as a
reduction of discrete particle noise and artificial two-body effects.  Due to
the similarity of collisionless dark matter systems and collisionless plasma
systems - both collisionless fluids governed by the Vlasov equation - the
phase-space sheet method may be readily applied to plasma simulations.  This
avenue is particularly well motivated as the phase-space sheet method has shown
improvements in exactly the main areas of weakness for the PIC method.  While
originally developed for cold dark matter fluids, we will demonstrate that the
method can be applied to arbitrary velocity distributions using an
initialization similar to the quiet start method.

The purpose of this paper is to apply the phase-space sheet method to the
simulation of collisionless plasmas, which we refer to as the simplex-in-cell
(SIC) technique.  In section \ref{sec:algorithm} we describe the SIC method and
its implementation in detail.  We benchmark the performance of the SIC method in
section \ref{sec:performance}, comparing it against PIC simulations on some
standard 1D electrostatic test problems. We discuss the results and give our conclusions in
section \ref{sec:discussion}.

\section{New Algorithm}\label{sec:algorithm}
\subsection{Description of New Method}

The classical \vp system is given by
\begin{equation}
\frac{\p f}{\p t} + \mathbf{v} \cdot \nabla_{\mathbf x} f 
- \nabla_{\mathbf{x}}\Phi \cdot \nabla_{\mathbf{v}} f = 0
\end{equation}
where $f=f(\mathbf{x},\mathbf{v})$ is the phase space density at the
six dimensional phase
space location $(\mathbf{x},\mathbf{v})$. The potential $\Phi$ is
given by the Poisson equation
\begin{equation}
\nabla^2\Phi = - C \rho(\mathbf{x}),
\end{equation}
which is sourced by  the charge density $\rho(\mathbf{x})$ and
related to the phase space density via the integral over all velocity
space coordinates 
\begin{equation}
\rho = \int f(\v{x},\v{v}) d^n v - 1.
\end{equation}
Here the mean density is set to zero when modeling situations with
periodic boundary conditions.
In an $N$-body method like PIC, this equation is discretized by writing the phase space
density as a sum of $N$ Dirac-$\delta$ functions (particles),
\begin{equation}
\tilde{f} = \sum^N_{i=1}  q_i
\delta(\mathbf{x}-\mathbf{x}_i)\delta(\mathbf{v}-\mathbf{v}_i),
\label{OldAnsatz}
\end{equation}
which gives a straightforward estimate of the charge density
\begin{equation}
\tilde{\rho}(\v{x}) = \sum^N_{i=1} q_i \delta(\v{x} - \mathbf{x}_i).
\end{equation}
We will use the tilde to denote approximate quantities throughout this paper.
 For the actual implementation, the particle quantities (such as charge and mass) are distributed onto a spatial grid using a shape function. The form of this shape function has been
explored widely, with triangular cloud approaches \cite{birdsall2004plasma}
 being perhaps the most common ones still in use today, as they are far less susceptible to grid heating than ``top hat''-shaped CIC particles.

The class of
solvers imagined by \cite{Abel:2012} and implemented in
\cite{Hahn:2012} differ fundamentally in that the ansatz~(equation
\ref{OldAnsatz}) is replaced. In a problem with $n$ spatial dimensions, instead of considering point particles, we will consider $M$ $n$-dimensional manifolds in $2n$-dimensional phase space. Assume that a given manifold, indexed by $j \in \{1,...,M\}$, is described by parametric functions $\v{x}_j(\v{l})$ and $\v{v}_j(\v{l})$, where $\v{l} \in [0,1]^n$ is a parametric coordinate that traces out the shape of the manifold in phase space\footnote{In one dimension for example, the point $(x_j(l),v_j(l))$ refers to the position in phase space of the point that lies a fraction $l$ along the manifold. The point $(x_j(1),v_j(1))$ would be the end of the manifold in phase space}. We can then write the contribution of the position $\v{l}$ on the $j$\textsuperscript{th} manifold to the total phase space density as
\begin{equation}
f_j(\v{x},\v{v},\v{l}) = \delta( \v{x} - \v{x}_j(\v{l})) \delta( \v{v} - \v{v}_j(\v{l})) \rho_j(\v{l})\;.
\end{equation}
Summing over all such manifolds, we obtain the contribution of all manifolds at the parametric point $\v{l}$ to the total phase space density:
\begin{equation}
f(\v{x},\v{v},\v{l}) = \sum_{j = 1}^{M} f_j(\v{x},\v{v},\v{l})\;.
\end{equation}
Intuitively, we have a set of points at positions $(\v{x}_j(\v{l}),\v{v}_j(\v{l}))$ with a given density $\rho_j(\v{l})$ for any given $\v{l}$. In order to obtain the contribution of the entire manifold, i.e. in order to obtain the actual phase space density, we need to integrate over $\v{l}$:
\begin{equation}
f(\v{x},\v{v}) = \int \sum_{j = 1}^{M} \delta( \v{x} - \v{x}_j(
\v{l})) \delta( \v{v} - \v{v}_j(\v{l})) \rho_j(\v{l})\; d^nl\;.
\end{equation}
The integral runs from $0$ to $1$ over all components of $\v{l}$.
The ``density'' $\rho_j(\v{l})$ is a density with respect to the parametric coordinate $\v{l}$. It satisfies
$$\int_{\v{l} = 0}^{1} \rho_j(
\v{l}) d^n l = q_j\;,$$
where $q_j$ is the total charge on the $j$\textsuperscript{th} manifold. It is also easy to see (by integrating first over $\v{x}$ and $\v{v}$ and then over $\v{l}$) that
$$\int_{\v{l}} \int_{\v{x}} \int_{\v{v}} f(\v{x},\v{v},
\v{l})_j \;d^nx\; d^nv\; d^nl = q_j\;,$$ as one would expect.

The phase space density is approximated by a finite number of sheets which are
infinitesimally thin. In the general six dimensional case this means $f$ is given by a sum over $M$ distinct three dimensional manifolds. All of those manifolds evolve in phase space and are in general themselves strongly deformed. The charge density now becomes 
\begin{equation}\label{eq:charge_density_l}
\rho(\v{x}) = \int_{\v{v}} f(\v{x},\v{v}) d\v{v} = 
\sum_{j = 1}^{M} \int_{\v{l} = 0}^{1} \delta( \v{x} - \v{x}_j(
\v{l})) \rho_j(\v{l})\; d^n l\;.
\end{equation}
In order to self-consistently evolve the manifolds, the key question now is how to best compute this charge density as a function of position, i.e. how to deposit the charge onto position space\footnote{Of course we can replace every occurrence of the word ``charge'' with ``mass'' in the above discussion.}.

Instead of allowing $\rho_j(\v{l})$, $\v{x}_j(\v{l})$ and $\v{v}_j(\v{l})$ to be any arbitrary functions of the continuous variable $\v{l}$, we now discretize and define $n_{particles/stream}^n$ tracers along the manifold (or ``stream''), indexed by a discrete variable $\v{i}$, and with fixed charges and phase space positions $q_j(\v{i})$, $\v{x}_j(\v{i})$ and $\v{v}_j(\v{i})$ ($\v{i} \in \{1,...,n_{particles/stream}\}^n$) for each tracer\footnote{In our notation, there are $n_{particles/stream}$ particles along every one of the $n$ dimensions of the manifold}. The functions $\rho_j(\v{l})$, $\v{x}_j(\v{l})$ and $\v{v}_j(\v{l})$ then become interpolating functions over the tracers\footnote{One can think of a mapping between the ordering of the tracers over $\v{i}$ and the corresponding position $\v{l}$ along the parametric manifold.}.The tracers and the interpolation scheme form an approximation to the idealized continuum quantities and allow us to approximate the computation of the charge density as in equation \ref{eq:charge_density_l}.

It is noteworthy to point out that PIC particles used in simulations are given by a $\delta$-function in velocity space and an extended shape function in position space. As such a particle is advected through phase space, its shape remains constant exactly (i.e. it will still be a $\delta$-function in velocity space with the same shape function in position space). By contrast, the particles between the tracers in SIC are a sub-manifold of the surrounding phase space and change their shape in both position and velocity space. For instance, in $1$D SIC, a particle is a line in phase space, which can stretch and rotate, while a PIC particle can merely advect in phase space but retains its rotational angle and shape.

\cite{Abel:2012} used the simplest possible interpolation for the
charge density (mass density in the dark matter case). They suppose
that given a set of phase space points on a three dimensional manifold
and a tessellation --- i.e. connectivity information of each
point with another 3 points defining a tetrahedron --- that the charge in
each tetrahedron $q_i$ is an invariant of the evolution. They consequently
assume a piecewise constant charge density of $q_i/V_i$ within the tetrahedron,
where $q_i$ and $V_i$ are simply the charge and
volume of the tetrahedron labeled by $i$. They were concerned
primarily with the 6 dimensional case. Of the natural analogs in one
and two dimensions we will discuss the one dimensional case in some
detail. This will allow us to develop the method in a setting where we can still draw and imagine the entire phase space structure
easily.

\subsection{Implications of the Sheet Topology}
We note that one of the important assumptions for our definitions of particles and tracers is the connectivity of the sheet in phase space: We defined a connected sheet in phase space as an ordered collection of tracers where a particle is defined by each pair of adjacent tracers. Even if tracers move far apart from each other or cross in position space. Further, we note that this method of a single sheet is easily extensible to several sheets, which allows the representation of more complex velocity distributions. The sheet topology in phase space has the consequence that collisions cannot be modeled as easily as with traditional PIC. Random collisional interactions between tracers would scatter the tracers into a continuous distribution of velocities, which is in conflict with the goal of discretizing phase space using connected sheets that maintain a fixed connectivity and small velocity differences between tracers.

The ordering of the tracers allows for an efficient translation of the algorithm into actual code. The linear layout of the tracers and the fixed, defined connectivity between adjacent tracers allows us to lay out connected tracers in \emph{physically} adjacent memory on the computer. In one dimension for example, this means that the \emph{scatter} step can be formulated as an $O(N)$ walk over an array of all particles, \emph{scattering} charge and current of one pair of adjacent tracers at a time onto the grid, without any searching and sorting. Several possible strategies may be fruitful for parallelizing this method. In 1D where it is reasonable to keep the entire simulation space in memory, every full sheet can be delegated to a processor. In higher dimensions, the simulation can be broken up into regions in phase space, with sheets broken up among those regions. ``Ghost tracers'' which give the positions of adjacent tracer particles in neighboring simulation regions would have to be communicated between processors to complete the deposit.

Another advantage of the sheet-topology is that the algorithm allows us to distinguish between phase space sheets. This means that at any point in time during the simulation, we can define exactly which sheet is located where, what the density is per sheet, how many times the sheet folded, how many sheets are at one point in position space, and so on. It further allows us to define a velocity distribution at \emph{any} point in space, not just at the grid points or tracer positions. For a given finite interval of arbitrary size in position space we interpolate the velocity of each particle that overlaps with that interval and perform a weighted sum over all overlapping particles\footnote{The procedure is exactly equivalent to that outlined in equations \ref{eq:overlap}, \ref{eq:rho_sum}, and \ref{eq:j_sum}, but the overlap $o_{ji}$ must now be taken over the interval in question}. Since we have fully connected sheets spanning our entire simulation region, we \emph{always} have a well-defined distribution function at any point in real space, and thus we have access to a well-defined velocity distribution (and similarly to other quantities such as the charge density, the current density, etc...) at \emph{any} point in the physical simulation region.

\subsection{1D Line Segments}\label{sec:segments}

Our initial algorithm is an adaption and variant of the approach presented in \cite{Abel:2012}. In this approach, the tracers and a tesselation among them denote only the \emph{boundaries} of the actual particles, which are defined between the tracers. These particles in turn are assumed to carry a fixed total charge with piecewise constant charge density. In our method, the tracers are being advected, while the charge is carried (and scattered) by the particles defined between them.

The key feature is thus to switch the spatial basis functions from Dirac-$\delta$ functions as in PIC to so-called ``box''-functions delineated by tracers. Given a pair of tracers $i$ and $j$, they define a one dimensional box function, $b_{i,j}(x,t)$ as
$$
b_{i,j}(x,t) = \left\{
\begin{array}{ll}
    n_{i,j} & \mbox{\rm{if}} \operatorname{min}(x_i(t),x_j(t)) < x < \operatorname{max} (x_i(t),x_j(t)) \\
    0 & \mbox{\rm{otherwise}}
\end{array}
\right. \; ,
$$
where
\begin{equation}\label{eq:piecewise_constant}
n_{i,j} = \frac{1}{|x_i(t) - x_j(t)|}\;
\end{equation}
is an effective local density of the particle defined by the spacing of the tracers\footnote{In practical computational applications, both $x_i(t)$ and $x_j(t)$ are float values and should not be equal. A suggestion would be to check for this case, and if so, randomly offset one of the particles by the minimal float value. As long as the total charge is retained, the exact handling of this case is not important, since the charge density is never referenced directly, but only integrated over (total charge is deposited onto the grid). Physically, we might imagine the minimal value of $x_i(t) - x_j(t)$ as the size of an actual plasma particle.}, and thus we have the normalization
$$
\int\limits_V b_{i,j}(x,t) dx = 1
$$
over the total simulation region $V$. In this definition, $x_i(t)$ and $x_j(t)$ are called the \emph{tracers} of the box function. This is the central idea of our approach: the spatial distribution function $n(x,t)$ in one dimension is defined through a connected set, or \emph{sheet}, of tracers and expanded as a set of box functions defined between \emph{adjacent} tracers. In higher dimensions, the connectivity gives rise not to a set of line segments but rather to a set of simplices. Each simplex then carries a charge and mass density. Formally, we thus write the spatial density $\rho(x,t)$ as
$$
n(x,t) = \sum\limits_{i=1}^{N} b_{i, i+1}(x,t) \; ,
$$
where for periodic boundary conditions, $x_{N+1} \equiv x_1$. Since we have now defined our particles to be the box functions, we must define a velocity for them in order to obtain the full distribution function $f(x,v,t)$. It is natural to define the velocity of the particle to be the weighted average of its tracers. Assuming that both tracers have an equal charge-to-mass ratio, we write
$$
f(x,v,t) = \sum\limits_{i=1}^{N} b_{i, i+1}(x,t) \delta(v-\tilde{v}_i) \;
$$
where
$$
\tilde{v}_i \equiv \frac{ m_i v_i + m_{i+1} v_{i+1} } { m_i + m_{i+1} }
$$
and $v_i$ and $m_i$ are the velocity and mass of the $i^{th}$ tracer, respectively.

The particles carry a fixed charge and a fixed mass, while the size and position (and thus the density) of the particles are defined by the tracers. Given the standard approach of discretization of the position space of the simulation into cells, this allows us to \emph{scatter} the charges and currents of the particles in a straightforward fashion. If we define the fractional overlap of the \( i^{th} \) particle with the \( j^{th} \) cell as \( o_{ji}\):
\begin{equation}\label{eq:overlap}
o_{ji} = \int\limits_{cell\;j} b_{i,i+1}(x) dx \; ,
\end{equation}  
we can write
\begin{equation}\label{eq:rho_sum}
\rho_j = \frac{1}{\tilde{V}}\sum\limits_{i}o_{ji} q_i
\end{equation}
for the charges and similarly
\begin{equation}\label{eq:j_sum}
\v{j}_j = \frac{1}{\tilde{V}}\sum\limits_{i}o_{ji} \v{I}_i
\end{equation}
for the current densities, where \( \v{I}_i \) is the current contribution from the \( i^{th} \) particle, defined by
$$
\v{I}_i = q_i \v{v}_i \; .
$$
Here \( \v{v}_i\) is the velocity of the \( i^{th} \) particle, \(\tilde{q}_i\) is the charge of the \(i^{th}\) particle and \(\tilde{V}\) is the volume of a cell.
The core of the algorithm thus amounts to finding the overlaps according to equation \ref{eq:overlap} and performing the sums \ref{eq:rho_sum} and \ref{eq:j_sum}. Because in a generic number of dimensions the quantities carried by a simplex are scattered onto the simulation region cells, we term the method ``simplex-in-cell''.

Apart from this modification to the \emph{scatter} step, the tracers can be treated as if they were particles in a PIC code. Using the charge density in the position grid, we can \emph{solve} for the electric field. We then interpolate (or \emph{gather}) the electric forces (which are well-defined on the spatial grid points) back onto the tracer particles. Using these forces and the tracer velocities, we then \emph{move} the particles, i.e. update their position and velocity. We see that the \emph{solve}, \emph{gather} and \emph{move} steps of a traditional PIC implementation \emph{remain unchanged}: we simply advect the tracers as we would the particles in the PIC code given their fixed charge-to-mass ratio. In practice then, we can reuse most of the existing algorithm \emph{without modification} for a given implementation of a simulation code; only the \emph{scatter} function must be rewritten. 

In our particular implementation, the forces are ``gathered'' onto the tracers using linear weighting (as in PIC with linear particles). Because the scatter and gather steps are not necessarily symmetric operations anymore, momentum conservation and the avoidance of unphysical self-forces are not guaranteed a priori as they are for PIC. However, we find in practice that momentum is conserved to the same degree of accuracy as it is for PIC.

\subsection{First-Order Weighting}\label{sec:first_order}
In a particle-mesh force calculation, such as that used in PIC, the simulation particles have continuous spatial 
coordinates while the field quantities are defined only at discrete grid points on a mesh.  
To construct a discrete charge density from the particles it is 
necessary to employ a weighting 
scheme that assigns their charges to the grid points on the mesh.
The simplest choice of weighting scheme is zero-order weighting, also known as nearest-grid-point (NGP).  Using zero-order 
weighting,
all of the charge from each particle is assigned to the grid point nearest that particle.  In the 1D case this
leads to the following discrete charge density:
$$\rho_{j} = \frac{1}{L}\int \sum_{i=1}^{N} q_{i} \, \delta(x - x_{i}) W_{0} (x - X_{j}) dx$$
where L is the cell length (equal to the spacing between grid points), the index $i$ labels the particles,
$W_{0}(x - X_{j})$ is the zero-order charge assignment function,
$$W_{0}(x - X_{j}) = \left \{ \begin{matrix*}[l]
1 & \rm{for} \left | x - X_{j} \right | < \frac{L}{2} \, or \, x - X_{j} = \frac{L}{2}\\ 
0 & \rm{otherwise}

\end{matrix*} \right.$$
\noindent and $X_{j}$ is the position of grid point $j$.

The zero-order scheme has the
advantage that the weighting step for each particle only requires accessing the data for a single grid point, making 
it computationally inexpensive.  However, as a particle passes through a cell boundary all of its charge is abruptly transferred 
from one
grid point to another, causing noise in the field quantities that can result in unphysical effects.  This problem can be reduced
through the use of higher-order weighting schemes, where the charge from each particle is distributed among multiple of its
nearby grid points.  This leads to  field quantities that are smoother in space and time, at the
expense of increased amounts of computation.  A commonly used higher order weighting scheme that may have a more acceptable 
balance between
computational cost and accuracy is first-order weighting, also known as cloud-in-cell (CIC).  With first-order weighting the
 charge of 
each particle
is linearly weighted to the $2^{n}$ grid points that lie on the corners of the n-dimensional cell that contains that particle.  
In the 1D case this gives

$$\rho_{j} = \frac{1}{L}\int \sum_{i=1}^{N} q_{i} \, \delta(x - x_{i}) W_{1} (x - X_{j}) dx$$
where $W_{1}(x - X_{j})$ is the first-order charge assignment function.
$$W_{1}(x - X_{j}) = \left \{ \begin{matrix*}[l]
1 - \frac{\left| x - X_{j} \right |}{L} & \rm{for} \left | x - X_{j} \right | \leq L\\
0 & \rm{otherwise}

\end{matrix*} \right.$$

Using first-order weighting, the discrete charge density values vary smoothly as a particle passes through a cell boundary, but the 
weighting step for each particle now requires 
accessing the data from two grid points, rather than just one as for zero-order weighting.

The method described by eq. (\ref{eq:rho_sum}) for defining a discrete charge density using the line segment interpretation of the
simulation particles can be viewed as a zero-order weighting of the charge from the line segments to the grid.  
With this view, each infinitesimal slice of a line segment is deposited to the grid according to zero-order 
weighting.  Analogous to PIC, first-order weighting can then be defined for line segments by depositing the infinitesimal slices of
each line segment to the grid according to first-order weighting.  The discrete charge density  obtained
in this way is
$$\rho_{j} = \frac{1}{L} \sum_{1}^{N} \int q_{i} b_{i, i + 1}(x) W_{1}(x-X_{j}) dx$$
where $b_{i,j}(x)$ is the box function defined in section \ref{sec:segments}.
Similarly, a first-order current density can be obtained by depositing the current from each infinitesimal slice of a line segment
to the grid 
according to first-order weighting.  The velocity of each slice is linearly interpolated from the velocities of the tracer 
particles that are the endpoints of the line segment.  In 1D, the current density at each grid point is then given by
$$j_{j} = \frac{1}{L} \sum_{1}^{N} \int q_{i} \tilde{v}_{i}(x) b_{i, i + 1} (x) W_{1}(x-X_{j}) dx$$
where $\tilde{v}_{i}(x)$ is the velocity linearly interpolated along the line segment indexed by i
$$
\tilde{v}_{i}(x) = \left\{\begin{matrix*}[l]
v_{i} + (v_{i+1} - v_{i}) \left ( \frac{ x - x_{i}}{x_{i+1} - x_{i}} \right ) & \rm{for} \; x_{i} \le x \le x_{i+1}\\
v_{i+1} + (v_{i} - v_{i+1}) \left ( \frac{ x - x_{i+1}}{x_{i} - x_{i+1}} \right ) & \rm{for} \; x_{i+1} \le x \le x_{i}\\
0 & \rm{otherwise}

\end{matrix*}\right.
$$
First-order weighting of the line segments may reduce the noise in the density 
and current fields as compared to zero-order weighting, but at the expense of increased amounts of computation.

\subsection{Piecewise Linear Segments}\label{sec:piecewise_linear}
In equation \ref{eq:piecewise_constant}, we took the ansatz of assigning a piecewise constant density to a particle defined by its adjacent tracers $i$ and $j$. A straightforward extension is to assume a piecewise \emph{linear} charge density for each particle. Extending the notation in section \ref{sec:segments}, let us define 
$$\bar{x}_{ij}(t) \equiv \frac{1}{2} (x_i(t) + x_j(t))$$
as the centroid of the particle defined by the tracers $i$ and $j$. We then write the charge density (which is now a function of position $x$) in a piecewise linear way:
\begin{equation}\label{eq:piecewise_linear}
\tilde{\rho}_{i,j}(x) = \tilde{\rho}^0_{i,j}  + (x - \bar{x}_{ij}(t)) \pd{\tilde{\rho}_{i,j}}{x}\;,
\end{equation}
where $\tilde{\rho}^0_{i,j}$ is the zeroth order charge density from equation \ref{eq:piecewise_constant}
$$ \tilde{\rho}^0_{i,j} = n_{i,j} \frac{1}{2}(q_i + q_j) \equiv \frac{\frac{1}{2}(q_i + q_j)}{|x_i(t) - x_j(t)|}\;,$$
and
$\pd{\tilde{\rho}_{i,j}}{x}$ is our estimate for the gradient of  $\tilde{\rho}_{i,j}(x)$.
It is easy to see that due to the symmetry of the linear term around the centroid $\bar{x}_{ij}(t)$, the integral over the entire particle of our piecewise linear charge density is equal to the integral of the zeroth order density  $\tilde{\rho}^0_{i,j}$, i.e. this new approach is still exactly charge conservative.
We estimate the gradient using a central difference. Given a set of $3$ adjacent, consecutive segments (bounded by the consecutive tracers from $i - 1$ to $i + 2$) with zeroth order densities $\tilde{\rho}^0_{i-1,i}$, $\tilde{\rho}^0_{i,i+1}$, and $\tilde{\rho}^0_{i+1,i+2}$; as well as the respective centroids $\bar{x}_{i-1,i}(t)$, $\bar{x}_{i,i+1}(t)$, and $\bar{x}_{i+1,i+2}(t)$; we estimate the gradient as
\begin{equation}\label{eq:gradient_rho_estimate}
\pd{\tilde{\rho}_{i,i+1}}{x} \equiv \frac{\tilde{\rho}^0_{i + 1,i+2} - \tilde{\rho}^0_{i - 1,i}}{\bar{x}_{i+1,i+2}(t) - \bar{x}_{i-1,i}(t)}\;.
\end{equation}
The deposit is then carried out as in section \ref{sec:segments}, but with the density from equation \ref{eq:piecewise_linear}. The piecewise linear estimate of the density is then an order higher in accuracy without sacrificing conservation or incurring a significant additional effort in the deposit.

There are cases where this estimation of the gradient in $\rho$ is problematic when the phase space sheet folds. In particular, it only makes sense if the ordering of the particles as shown in equation \ref{eq:gradient_rho_estimate} is also the positional ordering of the centroids of those particles, i.e. consecutively connected particles also have consecutive positions in real space ($\bar{x}_{i-1,i}(t) < \bar{x}_{i,i+1}(t) < \bar{x}_{i,i+1}(t)$ or the reverse). If the sheet folds, we might have a few cases at the edges of a fold where this is not the case, e.g. $\bar{x}_{i,i+1}(t)$ is larger (or smaller) than the positions of both of its neighbors. Estimating the gradient according to equation \ref{eq:gradient_rho_estimate} in this case does not make sense and could lead to large errors. We suggest having a simple check for this case in the code and simply not adding the gradient estimate when inappropriate. This should only affect a very small fraction of particles, since it is not the entire folded sheet itself that is problematic, but only the ``corner'', where a reversal of the ordering of the position of the particles takes place. As long as there are not too many such ``corners'' in the phase space sheet, this should not impact the performance of the first order weighting significantly, as it has not in our own tests.

\begin{figure}
\noindent\makebox[\linewidth]{
  \centering\subfloat[PIC]{
  \includegraphics[width=0.5\linewidth]{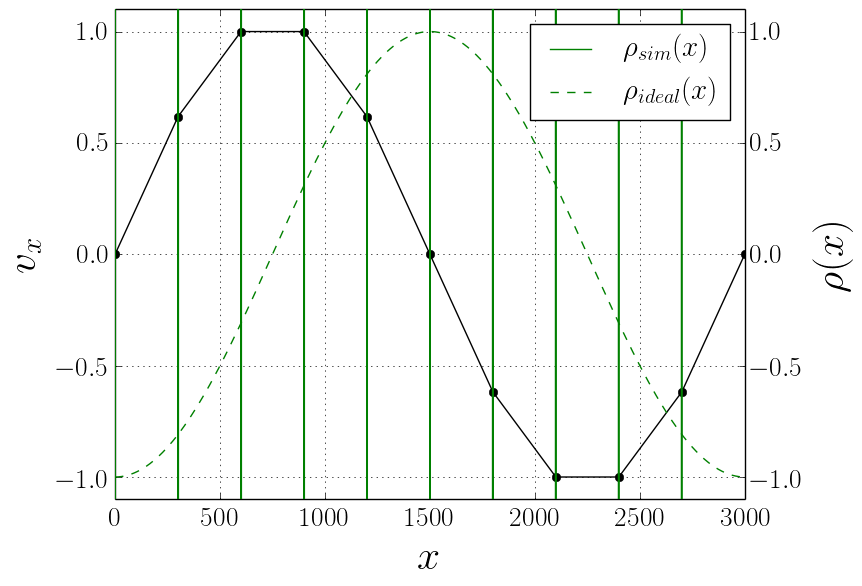}\label{fig:old_method_simple}
 } 
 }
\noindent\makebox[\linewidth]{
  \subfloat[SIC: piecewise constant segments]{
  \includegraphics[width=0.5\linewidth]{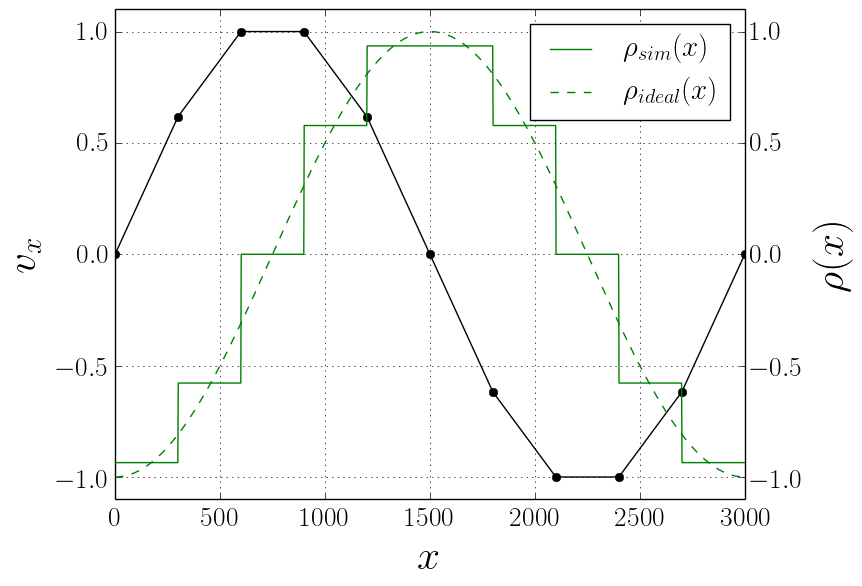}\label{fig:piecewise_constant_simple}
  }
  \subfloat[SIC: piecewise linear segments]{
  \includegraphics[width=0.5\linewidth]{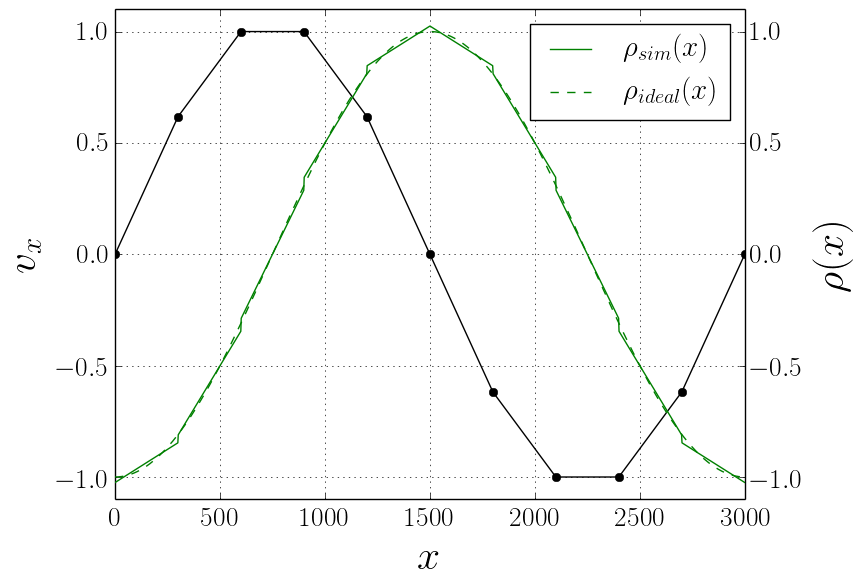}\label{fig:piecewise_linear_simple}
  }
}
  \caption{A comparison of the sheet location in phase space (black connected dots) and the resulting deposited charge density (green) for the old (PIC) method \protect\subref{fig:old_method_simple}, piecewise constant \protect\subref{fig:piecewise_constant_simple}, and piecewise linear segments \protect\subref{fig:piecewise_linear_simple}. Position is measured in units of the cell size $dx$ in this and all other figures. The test case is a cold, sinusoidally perturbed, electrostatic plasma in a periodic region where $n_{ppc} = \frac{10}{3000} \ll 1$. The charge density and velocity are normalized by their maximum value over the simulation region. It is clearly visible how the PIC particles produce very narrow peaks (limited in sharpness by the finite spatial grid). The functional form of the simulated charge density $\rho_{sim}(x)$ only begins to resemble the correct sinusoidal solution $\rho_{ideal}(x)$ once $n_{ppc}$ is set $\gg 1$. By contrast, both SIC schemes achieve a reasonable approximation of the sinusoidal density profile even though there are far fewer particles than cells. The piecewise constant \protect\subref{fig:piecewise_constant_simple} and piecewise linear segments \protect\subref{fig:piecewise_linear_simple} are clearly visible. The piecewise linear scheme visibly captures significantly more information by using gradient information.}
  \label{fig:order_comparison_simple}
\end{figure}

A comparison of PIC, SIC with piecewise constant charge densities, and SIC with piecewiese linear charge densities is given in figure \ref{fig:order_comparison_simple} with the example of a sinusoidally varying spatial charge density. We choose a relatively low particle number per grid cell, in order to emphasize the differences. Of course, as we increase the number of particles per cell, all solutions converge to the ideal sine wave solution.

\subsection{1D Refinement}
From the ansatz of assuming a piecewise constant or piecewise linear charge density between two tracers it becomes clear that the accuracy of the method is limited by the spacing of adjacent tracers. If two adjacent tracers move too far apart, then the information about the functional form of the charge density between the two tracers is only zeroth or first order over a wide range, which causes much information to be lost. 

For a simulation using the line segment method to accurately evolve the phase space sheet, the length of each individual line segment must thus be comparable to or less than the force resolution of the grid. There are two main cases in which this condition can be violated. The first and simplest case arises when the phase space sheet has a ``well-behaved'' topology\footnote{By this we mean that the sheet in phase space keeps its original order of tracers in position space, without folding or or stretching so as substantially change the positional spacing between the tracers}, but there are simply too few tracers. The second case arises when the sheet folds and stretches in phase space in such a way that adjacent tracers move far away from each other. A simulation of plasma instabilities such as the two-stream instability (as discussed in section \ref{subsec:twostream}) is an example of such complex behavior of the sheet in phase space.

In any situation in which the spacing between connected tracers becomes significantly larger than the grid spacing, it thus might become necessary to 
implement a refinement scheme in order to accurately continue evolving the system in time. One simple scheme is to have 
a condition on the maximum segment length (which will depend on the grid spacing). If at any time any segment becomes larger than 
the maximum length, a new tracer is inserted in the center of the segment (such that the two new segments each carry half of the 
charge of the old segment).  The velocity of the new tracer is initialized to the average of the velocities of the original two 
tracers.  More accurate refinement schemes may also take into account the distance between the tracers in velocity space or the
curvature of the phase-space sheet.  The simple refinement criterion we have implemented is sufficient for the initial 
demonstration of the SIC method, and more complex schemes will be investigated in future work.

\begin{figure}[!h]
\noindent\makebox[\linewidth]{
  \subfloat[]{
  \centering\includegraphics[width=0.5\linewidth]{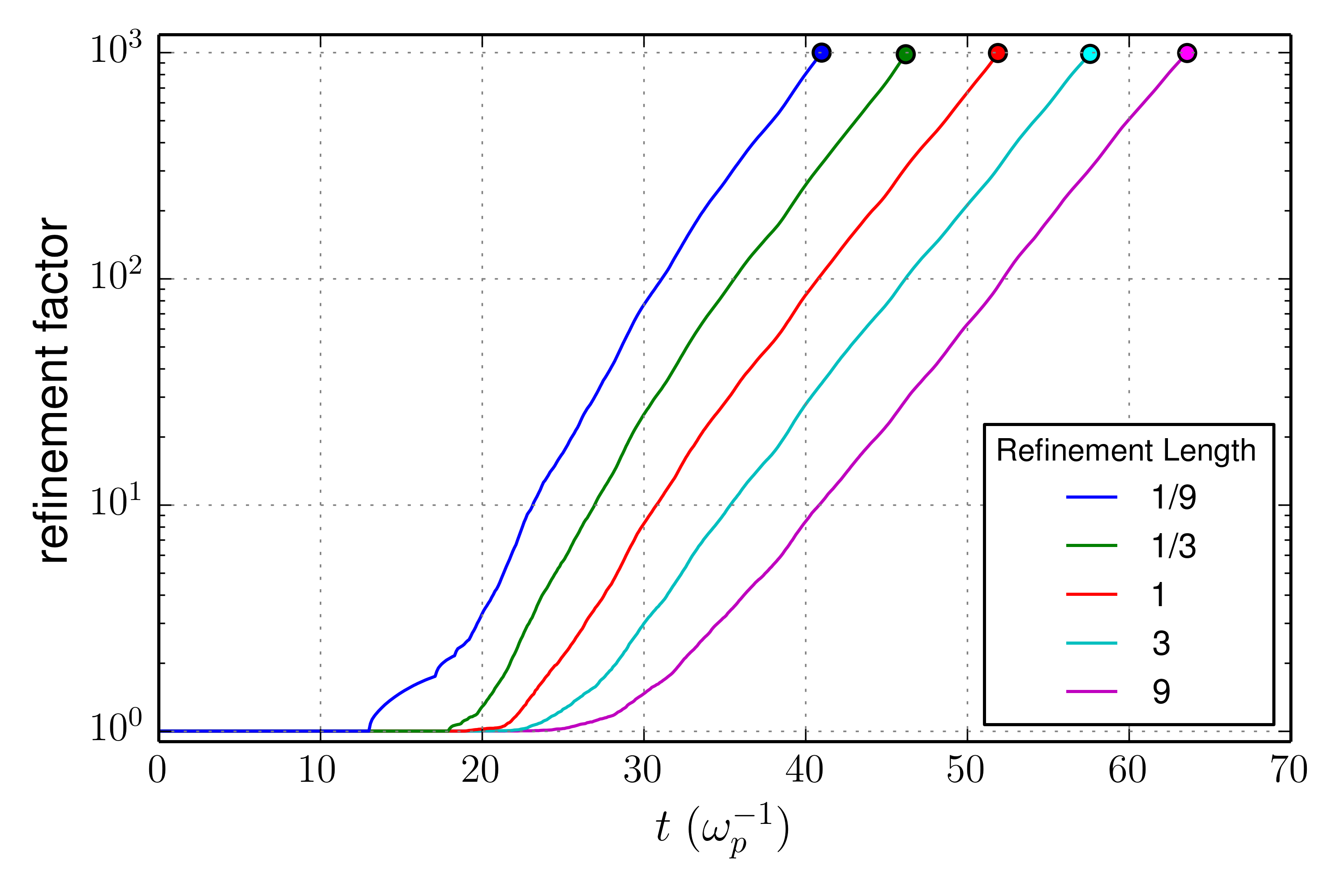}\label{fig:refinement_factor}
  }
  }
\noindent\makebox[\linewidth]{
  \subfloat[]{
  \includegraphics[width=0.5\linewidth]{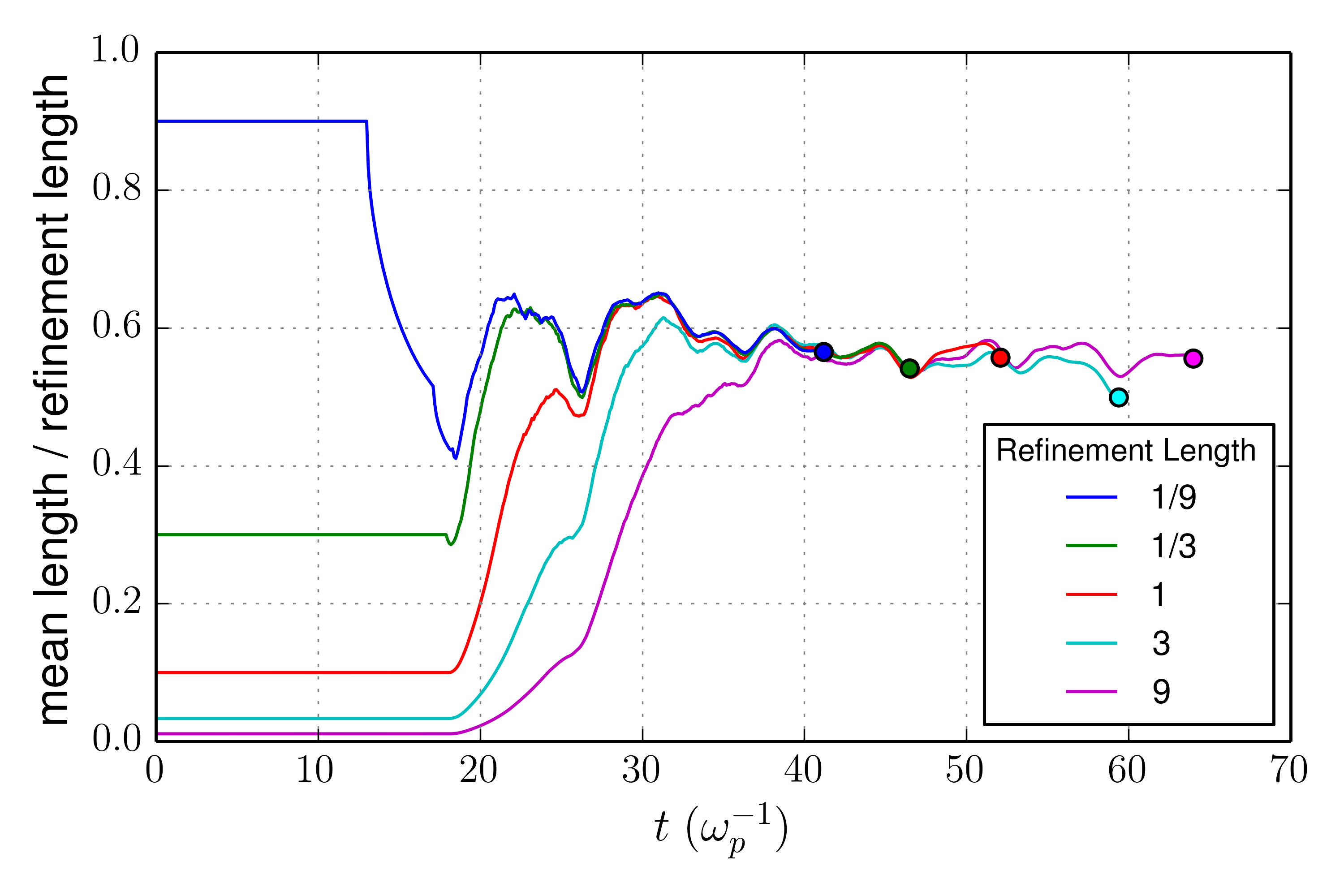}\label{fig:mean_length}
  }
  \subfloat[]{
  \includegraphics[width=0.5\linewidth]{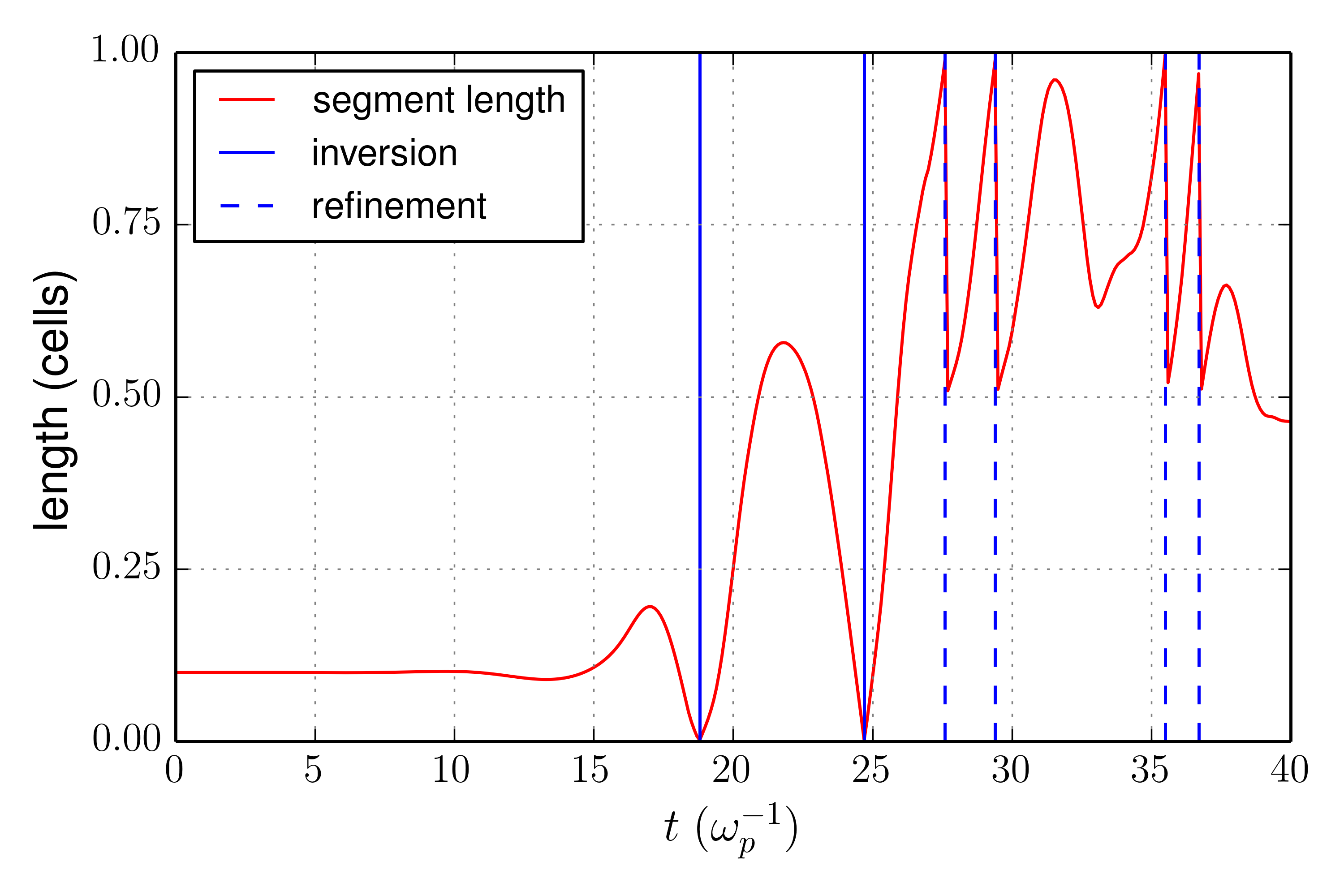}\label{fig:single_segment_length}
  }
}

  \caption{\protect\subref{fig:refinement_factor} Evolution of the refinement factor (current number of tracer particles / initial number of tracer particles.)
	 In \protect\subref{fig:refinement_factor} and \protect\subref{fig:mean_length} the circles
	 denote the point at which each simulation was stopped. \protect\subref{fig:mean_length} Mean line
	 segment length throughout the simulations, normalized to the refinement length. \protect\subref{fig:single_segment_length} Evolution of a
	 line segment during a simulation with a refinement length of 1 cell length.  Inversions mark when the endpoints of the
	 line
	 segments cross each other in position space, and refinements mark when the line segment has grown to be larger than the
         refinement length and is split in two.}
  \label{fig:refinement}
\end{figure}

Figure \ref{fig:refinement} demonstrates this refinement scheme for five different simulations of the two-stream instability.
The simulations all start with 10 particles per cell per stream\footnote{We perturb two anti-propagating cold streams with a small sinusoidal position perturbation at the fundamental mode of the box of amplitude $0.01 dx$ (where $dx$ is the grid spacing). The stream velocities are $\pm v_0$, where $v_0$ is chosen to maximize the growth rate of the perturbation (see section \ref{subsec:twostream}). We use $n_{cells} = 100$ and a timestep $dt = 0.1 \omega_p^{-1}$.}, but differ in the maximum segment length (referred to as the ``refinement 
length''.)  Each simulation was evolved until the refinement factor (current number of tracer particles / initial number of tracer
particles) reached $10^{3}$.  The refinement factor for each simulation is plotted in figure \ref{fig:refinement_factor}.  During the
linear growth phase (which saturates around $t \approx 17.5 \, \omega_{p}^{-1}$), the stretching of the line segments is small and 
there is
seen to be little refinement.  In the nonlinear regime the streams become severely distorted as a vortex is formed in phase-space.
Once this complex structure has begun to develop, the refinement factor grows at an approximately 
exponential rate\footnote{The exponential dependence arises because of repeated folding (filamentation) of the sheet in phase space.}.  The
refinement length controls the time of the onset of exponential growth, but has little effect on the growth rate.  

The evolution 
of the mean line segment length for each simulation, normalized to the refinement length, is plotted in figure \ref{fig:mean_length}.
The mean length tends to settle near a value slightly larger than one half of the refinement length for the simulations performed.  
For
the simulation with a refinement length of $1/9$ cells, the mean length initially decreases because its final value is less 
than
the initial particle separation.  For the other four simulations the initial particle separation is less than the final mean length,
and the mean length is seen to grow from its initial value.  The evolution of the length of a representative line segment is
shown in figure \ref{fig:single_segment_length}.  This illustrates two important types of events that are characteristic of the 
evolution
of a line segment.  During ``inversions'' the line segment length instantaneously becomes zero, which occurs when the tracer
particles that are the endpoints of that line segment cross each other and change their ordering along position space.
``Refinements'' denote the points in time at which the segment length has grown to be larger than the refinement length, and the
line segment is split in two.  The evolution plot shown consistently follows the portion of the original line segment
associated with a single one of its initial tracer particles.

\section{1D Neutral Plasma Tests}\label{sec:performance}
We now employ a set of standard \cite{rossmanith2011positivity,banks2010new,filbet2003comparison} plasma tests that have been used by several authors to evaluate new codes for the Vlasov-Poisson system. In particular, we use test problems for both warm and cold plasmas. We compare our own implementation of the SIC method with a standard PIC code. Both codes are identical, except for the \emph{scatter} step (see section \ref{sec:segments}), which is performed according to the scheme in question.

In one dimension, the simplex is a line segment. Thus, the $M$ manifolds
with which we discretize the initial phase space density are one
dimensional lines moving in the associated two dimensional phase
space. For a cold plasma (zero temperature) we can in fact start with
the single sheet case ($M=1$) before moving to cases with $M>1$. 
In fact, let us start with a uniform medium of charge density
everywhere and the cold initial phase space distribution 
\begin{equation}
f(x,v) = \rho_0(x)  \delta \left(v - v_s - v_1 \sin(2\pi x / L) \right)\label{coldSheet}
\end{equation}
where $L$ is the length of the simulation region.

\subsection{Plasma Oscillations of a Cold Sheet}

\begin{figure}[!h]
  \noindent\makebox[\linewidth]{
  \subfloat[Error Convergence as a function of $n_{ppc}$ ($n_{cells} = 389$).]{
  \includegraphics[width=0.52\linewidth]{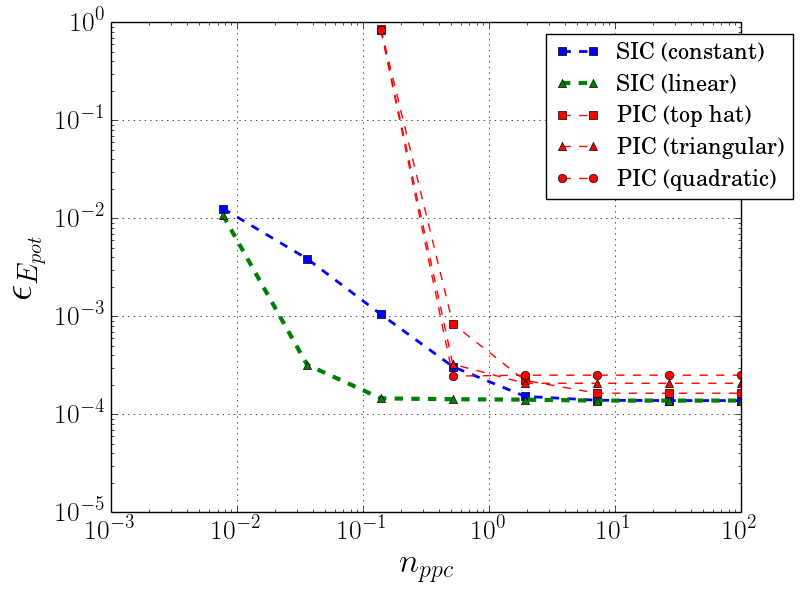}
  \label{fig:e_pot_error_simple}
  }
  \subfloat[Error Convergence as a function of $n_{cells}$ ($n_{ppc} = 100$)]{
  \includegraphics[width=0.48\linewidth]{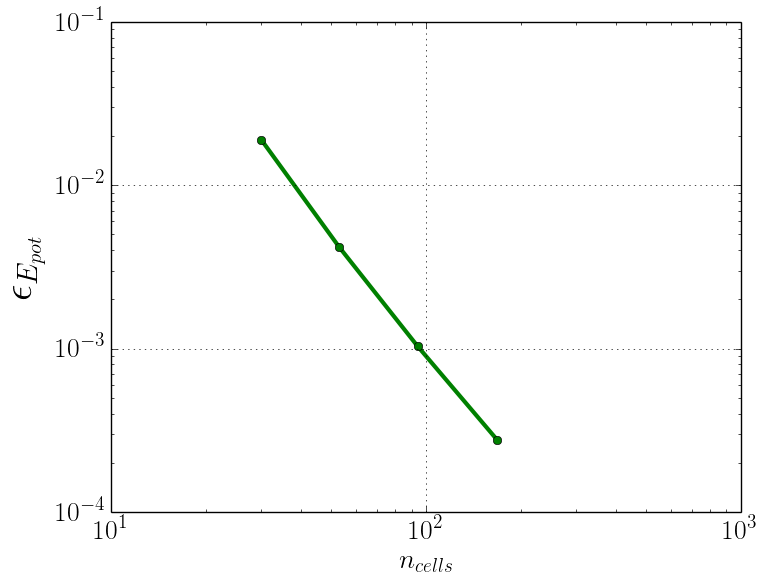}
  \label{fig:e_pot_error_simple_cells}
  }
  
}

  \caption{ \protect\subref{fig:e_pot_error_simple} The error in the potential energy evolution $\epsilon_{E_{pot}}$ for the old (PIC) and new (SIC) schemes as a function of particle number per cell. We employ CIC (``top hat''), triangular (``triangular'') and ``quadratic'' (i.e. cubic interpolation) particle shapes for PIC. We compare against SIC with the two cases of piecewise linear and piecewise constant charge densities on the segments. Both PIC and SIC converge to nearly the same ``base-line error'', set by when the error due to the finite grid spacing becomes dominant. However, SIC more rapidly reaches this baseline, especially with linear segments, which offer higher order convergence. Higher order particle shapes for PIC offer slightly lower errors than standard CIC (``top hat'') but converge to a higher baseline error due to their larger size. In \protect\subref{fig:e_pot_error_simple_cells}, we show the common baseline error reached by PIC and SIC as a function of the number of cells. The error was measured as the mean $L_1$ difference over one plasma period between $\tilde{E}_{\rm{pot}}(t)$ as obtained from the simulation and the analytical solution $E_{\rm{pot}}(t) = \sin^2(\frac{2 \pi t}{T_p})E_{tot}$, where $T_p$ is the analytical value for the plasma period.}
  \label{fig:errors}
\end{figure}

We apply the above described density assignment algorithm to the test case of electron plasma oscillations in a 1D plasma with periodic boundary conditions. We assume that the immobile ions produce a uniform positive charge background. We take $v_s=0$ and $v_1$ in equation
(\ref{coldSheet}) of some small value to study the evolution of plasma waves in our system.  This corresponds to a uniform space charge density and a sinusoidal perturbation of magnitude \( v_1 \) to the velocities. In that case, we expect simple standing electrostatic oscillations. In particular, the velocity distribution in phase space, \( v(x,t) \), will analytically have the form
$$
v(x,t) = v_1 \sin(2 \pi x / L) \cos( \omega_{p} t) \; ,
$$
where \( \omega_{p} \) is the plasma frequency.
    We can now compare both SIC and PIC to the analytical solution. The results are summarized in figure \ref{fig:errors}, where we study the error in potential energy\footnote{We set up a cold plasma oscillation of a single sheet by perturbing a stationary sheet with a sinusoidal velocity component at the fundamental mode of the simulation region. The perturbation amplitude is taken to be much smaller than the phase velocity of the resulting plasma oscillation $\frac{\omega_p}{2\pi/L}$}. The new scheme performs much better than PIC at lower particle numbers, while both schemes approach the same base-line performance as the particle number becomes very high, confirming consistency. At high particle numbers, the error baseline is due to the grid discretization of the simulation region into cells. This baseline error of course drops as the number of cells is increased.

    While this and the following tests focus mainly on the potential energy evolution $E_{pot}(t)$, this is just one of many possible testing grounds for the algorithm that we picked out for clarity and consistency. In this and the following investigations, we also monitored our algorithms for conservation of total momentum and energy. While SIC is neither explicitly momentum conserving (due to the lack of symmetry between the charge scattering and the force gathering steps) nor energy conserving, we find in all tests that both total energy and total momentum are conserved very well and always at least as well as in PIC with the same number of particles.

   We also studied the computational cost of PIC and SIC in all of our test problems. While these cases may not be entirely representative of large-scale, state of the art computational plasma physics tasks, we find as expected that PIC and SIC require nearly the same computational time for given values of $n_{ppc} \gtrsim 1$ and $n_{cells}$. PIC runs faster for $n_{ppc} << 1$ (because SIC requires touching all cells, even in this regime, while PIC only touches a small constant number for each particle), but in this regime PIC is highly inaccurate even for the simplest test cases.

\begin{figure}[!h]
  \centering\includegraphics[width=0.6\linewidth]{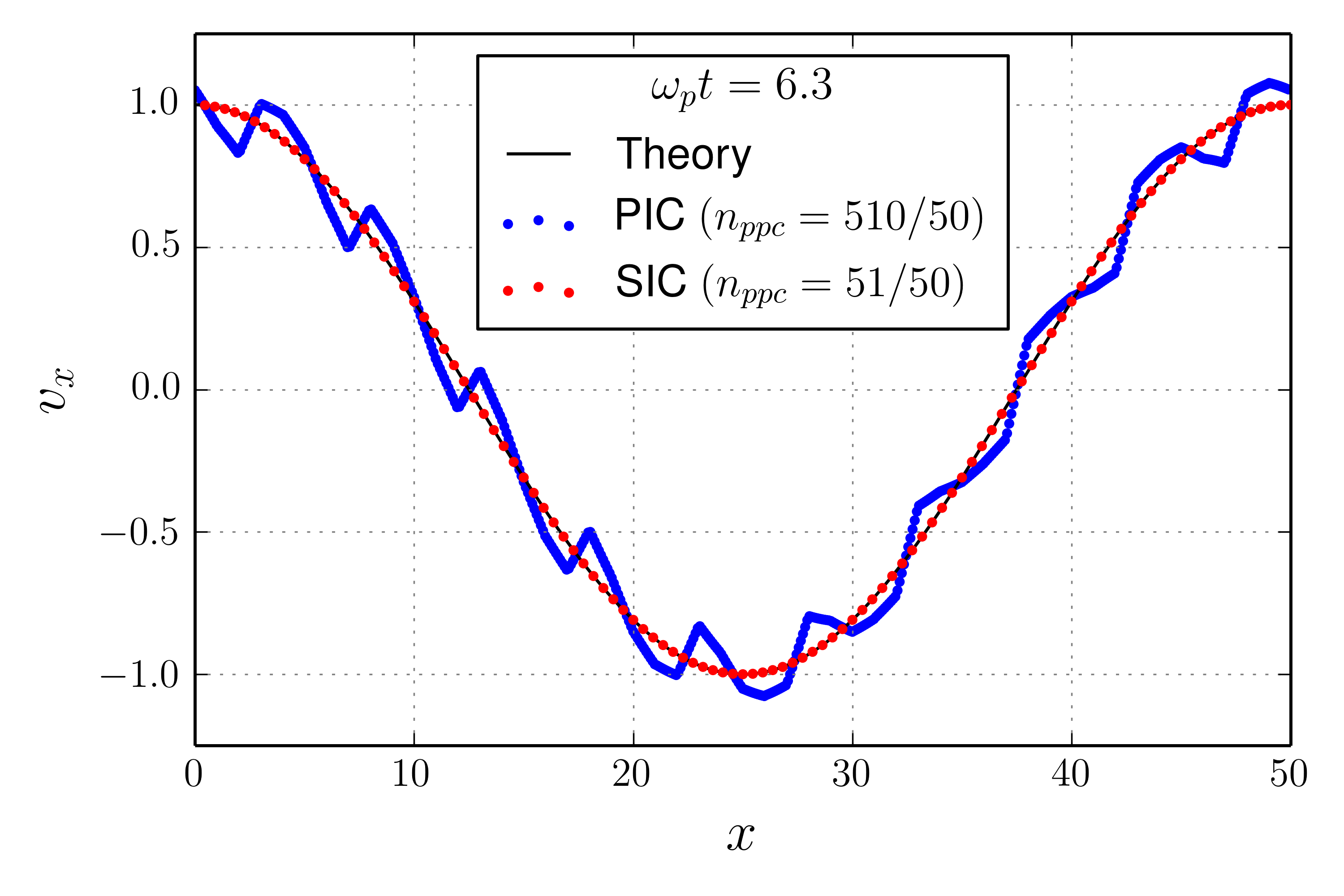}
  \caption{Comparison of phase-space for PIC with linear particles (``top hat'' shape) and SIC with constant segments after one plasma oscillation period.  The velocity scale is in units
           of the initial velocity perturbation amplitude $v_1$.  The grid induced noise seen when using PIC is absent in the
           simulation performed with SIC, even with $1/10$ the number of particles.}
  \label{fig:wave_comparison}
\end{figure}

Figure \ref{fig:wave_comparison} shows a comparison of phase-space after one plasma oscillation for two simulations, one using PIC
with 510 particles per 50 cells and the other using SIC with 51 particles per 50 cells.  These choices of parameters result in the
initial number of particles per cell not being uniform across the grid.  In the PIC simulation, significant noise from grid heating \cite{langdon1970effects} has developed
in phase-space after only one plasma oscillation period. This noise is eliminated when the system is evolved using SIC,
even with one-tenth the number of particles. We used linear particles (``top hat'' shape) for PIC. For higher-order particles the noise is less pronounced, but it still arises if we decrease the number of particles to near $n_{ppc} = 1$. SIC remains without noise even in this regime. While SIC (because of its linear interpolation) may theoretically be subject to the grid heating instability just like PIC is, we find in our tests that SIC is much less exposed to this problem.

\subsection{Two-Stream Instability}\label{subsec:twostream}
As a classic test problem that exhibits nonlinear evolution of the distribution function, we perform simulations of the 
two-stream instability and compare the SIC and PIC methods to each other and to theory \cite{bittencourt2004fundamentals}
.  Linearizing the system of equations 
describing two cold plasma
streams with equal densities and opposite velocities $v_{0}$ in a uniform background of ions gives the following 
dielectric function:
$$\frac{1}{\epsilon_{0}} \epsilon(\omega, k) = 1 - \frac{\omega^{2}_{p}}{(\omega + k v_{0})^{2}} - \frac{\omega^{2}_{p}}{(\omega - k  v_{0})^{2}}$$
For electrostatic (longitudinal) oscillations, setting $\epsilon(\omega, k) = 0$ gives a quartic equation for $\omega$ that leads to
four dispersion branches, with unstable modes existing for
$$k < \frac{\sqrt{2}\omega_{p}}{v_{0}}$$
In our simulations the distribution function is initialized as
$$f(x,v) = \delta(v \pm v_{0}) (1 + \epsilon \cos(kx))$$
where the perturbation wave vector $k$ is chosen to excite the mode with the maximum growth rate. According to the dispersion relation, this mode will have a growth rate of
$$\operatorname{Im}[\omega_{max}] = \frac{\omega_{p}}{2}\;.$$
The size of the simulation box is chosen such that the wave vector in question also corresponds to the fundamental mode of the box.

\begin{figure}[!h]
  \noindent\makebox[\linewidth]{
    \subfloat[Energy in the excited mode of the electric field]{
      \includegraphics[width=0.5\linewidth]{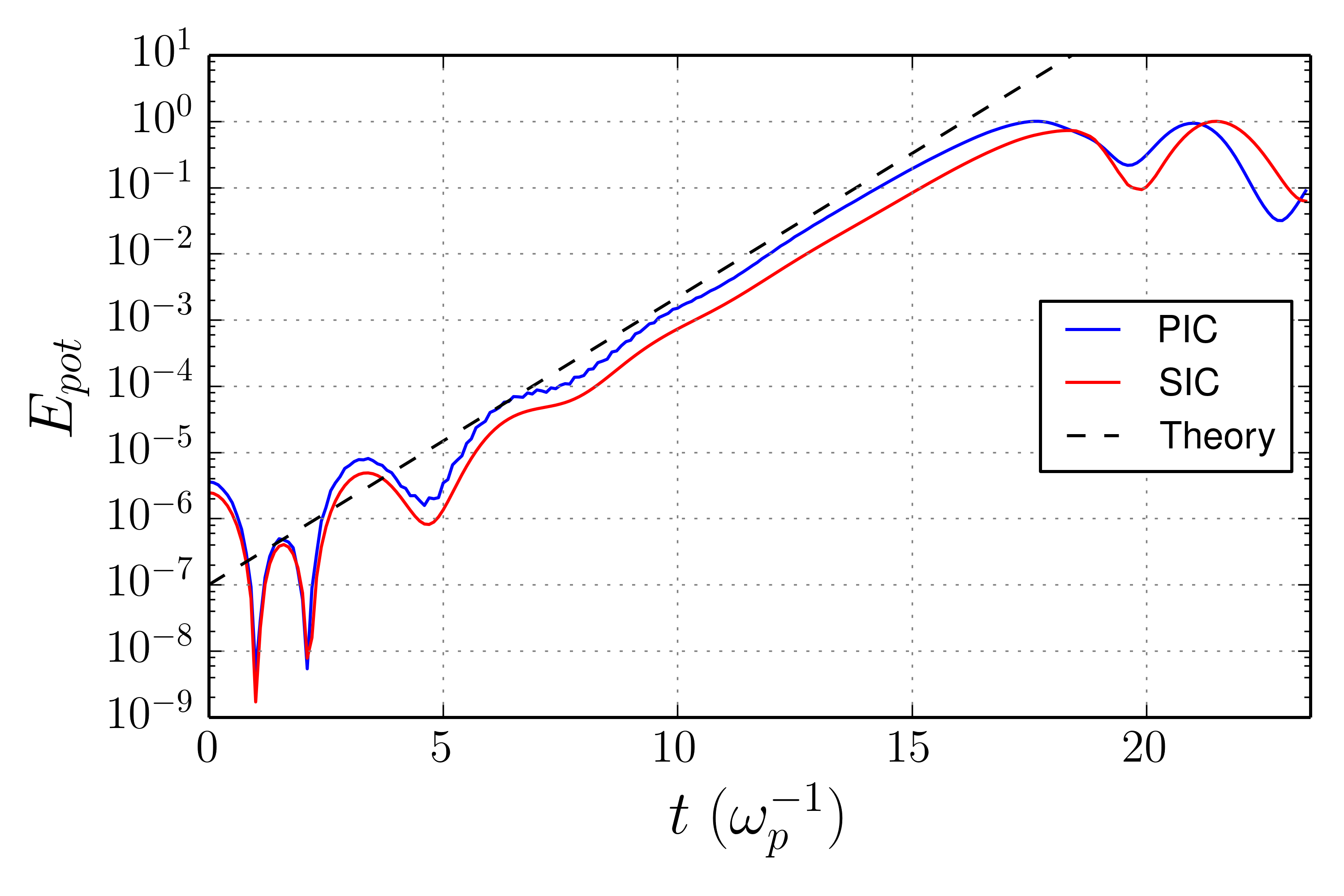}\label{fig:E_total_energy}
    }
    \subfloat[Total electric field energy]{
      \includegraphics[width=0.5\linewidth]{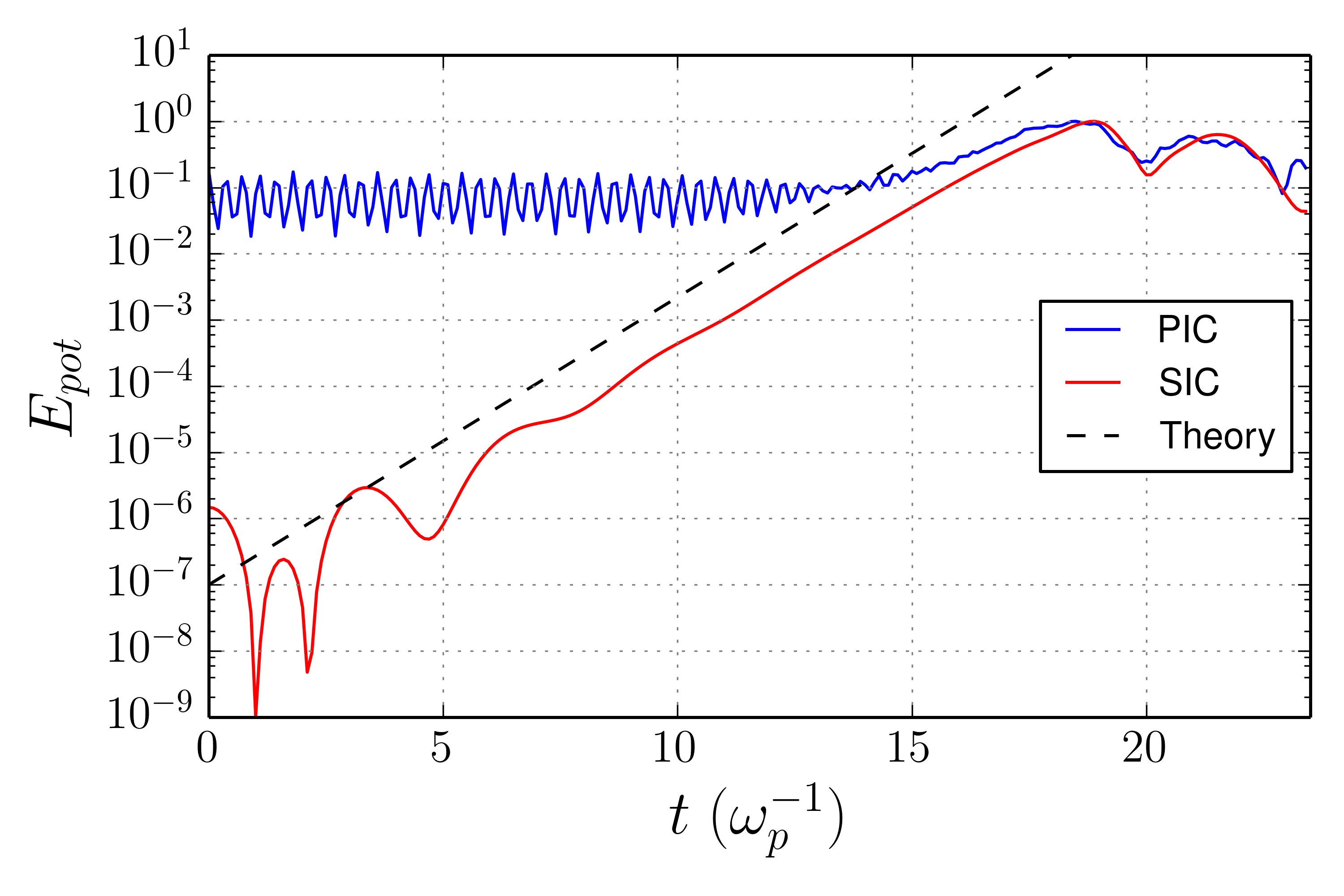}\label{fig:E_mode_energy}
    }
  }  
  \caption{Evolution of the energy in the excited component of the electric field and the total energy in the electric field during a two-stream instability. The simulations have $0.1$ particles per cell per stream, for both SIC and PIC. Although both methods accurately reproduce the exponential growth of the excited mode, the noise introduced by PIC excites many other spurious modes, and thus only SIC is able to correctly reproduce the growth of the total electric field energy.}
  \label{fig:E_energy_comparison}
\end{figure}

\begin{figure}[!h]
  \centering\includegraphics[width=0.6\linewidth]{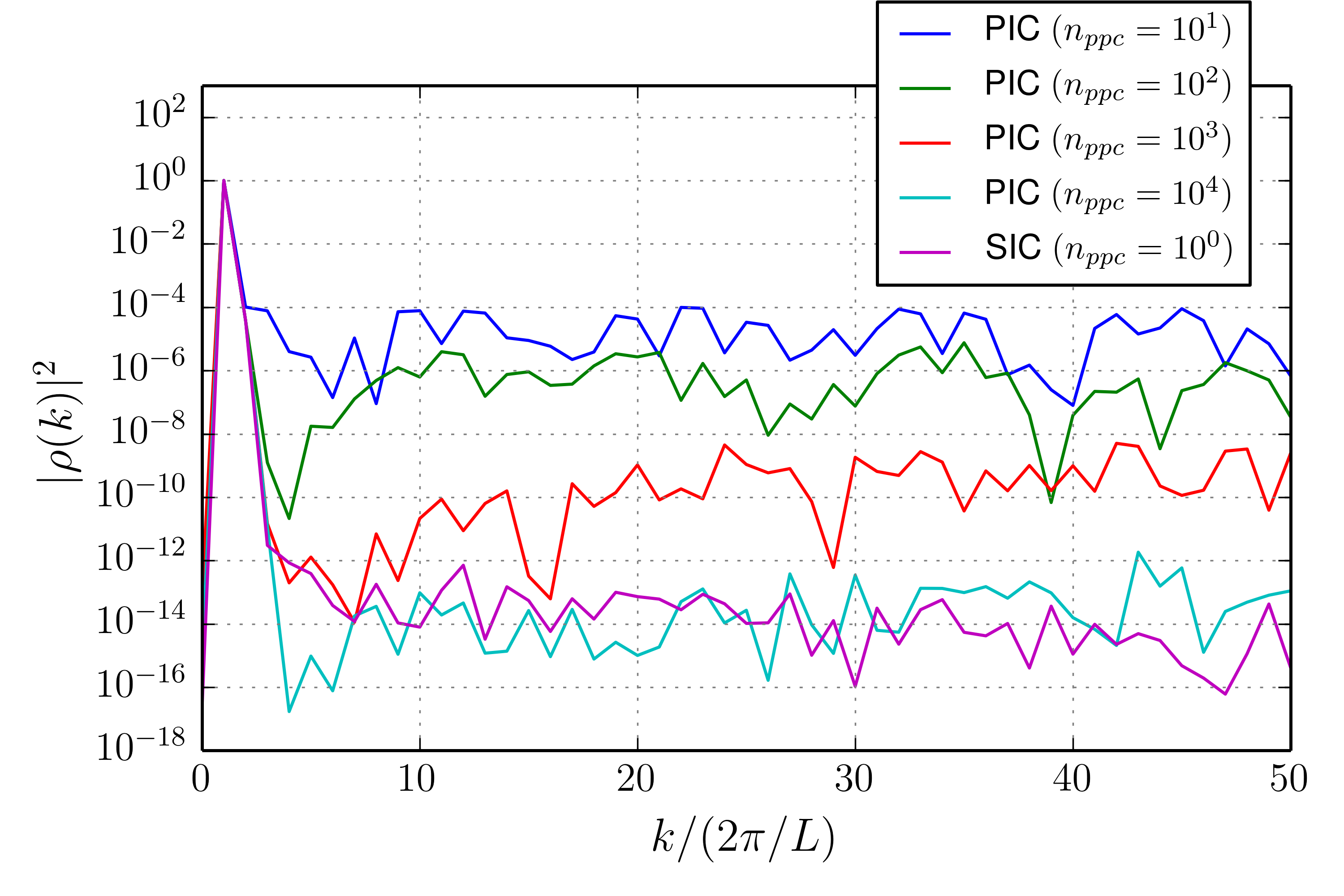}
  \caption{Fourier transform of the charge density in the simulation of the two-stream instability. The charge density is a snapshot in time during the linear growth phase of the instability. Note that in addition to the explicitly excited mode, there is significant noise in many other higher modes for PIC. Only with high particle numbers does the noise level approach that of SIC.}
  \label{fig:ts_fourier}
\end{figure}

Figure \ref{fig:E_energy_comparison} shows a comparison of the time evolution of the electric field energy for the two methods, 
for simulations with $0.1$ particles per cell per stream\footnote{The parameters are the same as for figure \ref{fig:refinement}, but with $n_{ppc} = 0.1$ instead of $10$.}. Both methods are able to accurately capture the exponential growth of the
excited component of the electric field, giving a rate that closely agrees with the analytical prediction.  However, only the SIC
method is able to correctly reproduce the evolution of the total electric field energy.  When using the PIC method, such low
particle numbers result in a very noisy charge density and resulting electric field, exciting many modes other than just the
fundamental that was explicitly excited in the initialization of the distribution function. The presence of these higher
modes in the electric field will cause effects on small spatial scales that do not accurately model the intended distribution
function, where these modes are not present.  Whenever such spatial scales are of interest, the SIC method has the clear
advantage that it is able to accurately evolve the distribution function at particle numbers much lower than possible using
the PIC method, reducing the computational cost of the simulation.  This point is also confirmed figure \ref{fig:ts_fourier}. Only with much higher number of particles is the lower-wavelength noise removed when using PIC.

The SIC method is able to accurately simulate the initial linear growth of the instability without refinement of the line
segments, (as shown in figure \ref{fig:E_energy_comparison} where no refinement was used for the SIC method).  However, after
the linear growth saturates the line segments begin to rapidly expand as a vortex develops in phases space that continuously
stretches the phase space sheets of the two streams.  Once a single line segment has expanded to the size of the simulation
box, the simulation can no longer continue as it is not possible to contain a line segment larger than the periodic simulation
region.  This problem is overcome by refining the line segments based on a refinement length criteria, allowing the simulation
to proceed into the nonlinear regime.  As shown in figure \ref{fig:refinement}, when using refinement to simulate into the nonlinear
regime, the number of line segments grows approximately exponentially with time.  The length of time that the simulation is
able to proceed is then limited only by the available computational resources.  While the exponential growth in the number of
particles means that the SIC method will be more expensive computationally than the PIC method when applied to the two-stream
instability, the SIC method retains the intricate phase space structure that develops as the vortex wraps and folds the
phase space sheets of the two streams into each other.  This information is completely lost when using the PIC method, likely
reducing the accuracy of simulations in the nonlinear regime.

\subsection{Landau Damping}

\begin{figure}[!h]
   
  \noindent\makebox[\linewidth]{
  \subfloat[Error Convergence as a function of $n_{ppc}$ ($n_{cells} = 389$)]{
  \includegraphics[width=0.52\linewidth]{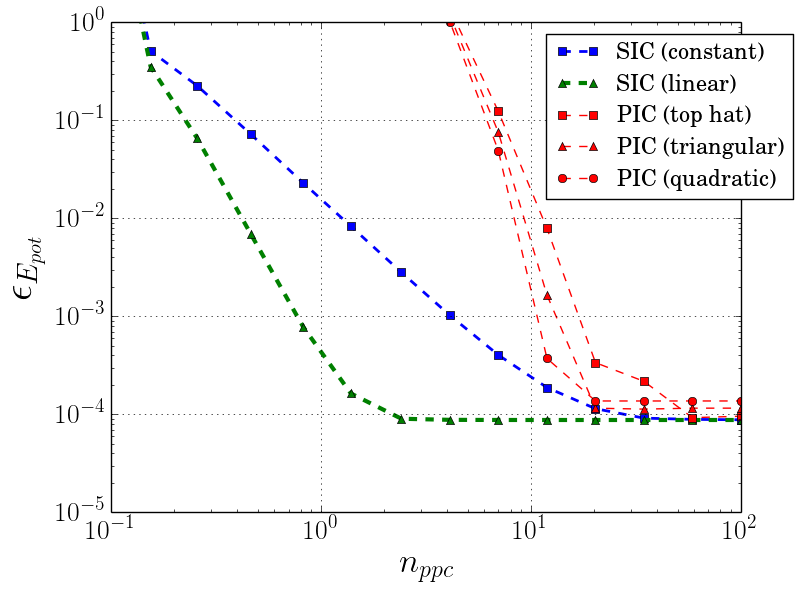}
  \label{fig:e_pot_error_landau}
  }
  \subfloat[Error Convergence as a function of $n_{cells}$ ($n_{ppc} = 100$)]{
  \includegraphics[width=0.48\linewidth]{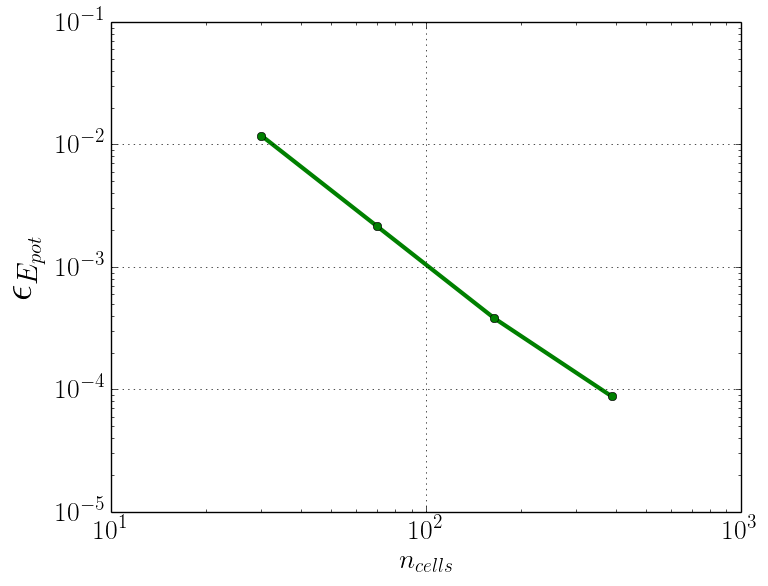}
  \label{fig:baseline_error_landau}
  }
  
}

  \caption{ \protect\subref{fig:e_pot_error_landau} Error convergence comparing the SIC method to the standard PIC method in the Landau damping problem. The figure has the same style as figure \ref{fig:errors}. Note how the new scheme converges faster, requiring far fewer particles than the old scheme for equivalent errors. Note also the increase in convergence order when moving from piecewise constant to piecewise linear segments. A minimum error is reached by all schemes as in figure \ref{fig:errors} due to the discretization of the simulation region into cells. As the number of cells is increased, this ``baseline error'' decreases. The baseline error is plotted in \protect\subref{fig:baseline_error_landau} as a function of the number of cells. As the number of cells is increased further, the error again reaches a minimum (around $n_{cells} \sim 750$) due to the finite number of streams and the finite timestep. The convergence was obtained by comparing the evolution of $E_{\rm{pot}}(t)$ during the simulation to a ``gold'' run with very high particle number. The error represents the mean $L_1$ difference between $E_{\rm{pot}}(t)$ during a given run and $E_{\rm{pot}}(t)$ for the gold run. In this case, the gold run was performed with the old (PIC) scheme. $M = 20$ streams were used. These results show that in this problem SIC can outperform PIC in accuracy with far fewer resources.}
  \label{fig:ld_e_pot_error}
\end{figure}

\begin{figure}[!h]
\noindent\makebox[\linewidth]{
  \subfloat[horizontal streams]{
  \includegraphics[width=0.5\linewidth]{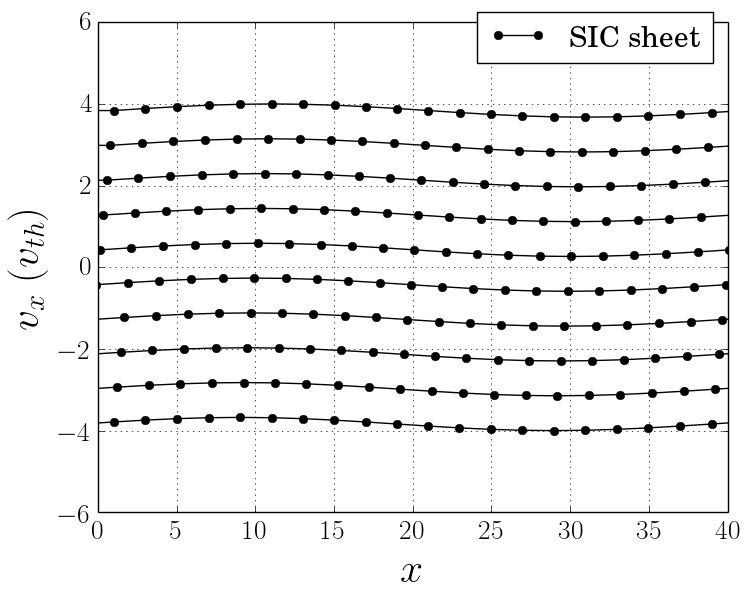}\label{fig:phase_space_setup_horizontal}
  }
  \subfloat[vertical streams]{
  \includegraphics[width=0.5\linewidth]{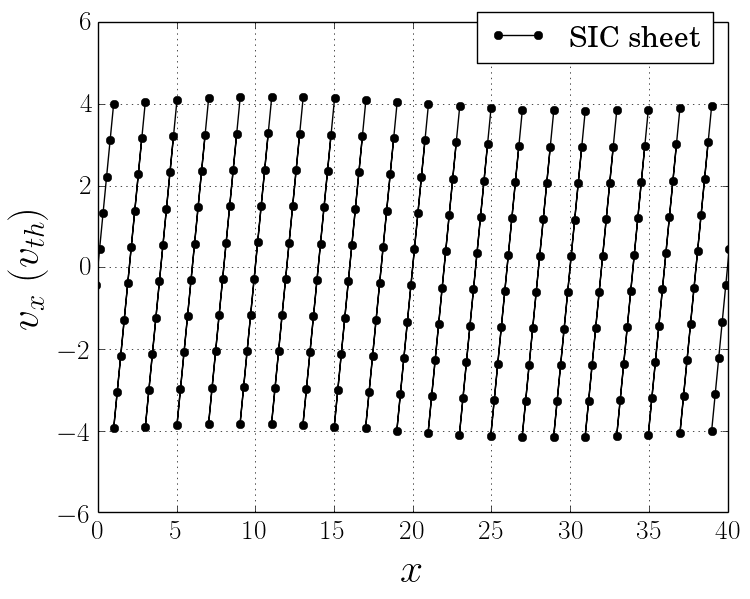}\label{fig:phase_space_setup_vertical}
  }
}

  \caption{A comparison of the geometry and connectivity of the SIC sheets in phase space for the case of ``horizontal'' and ``vertical'' streams. Velocity is measured in units of the thermal velocity. The phase space geometry represents a sinusoidally perturbed warm plasma with $M = 10$ streams and $n_{ppc} = 200$ particles per cell (distributed over the streams). The positions of the tracers in phase space are identical in both cases, but the connectivities are different: in the horizontal case, we have $10$ sheets with $20$ particles in each of them. In the vertical case, we have essentially $20$ vertical sheets, with $10$ particles in each of them. Both are shown one timestep after the beginning of the simulation. With periodic boundaries the connectivity of course ``wraps around'' the simulation region. As the simulation progresses, the horizontal streams advect and retain their shape approximately, while the vertical streams tilt to the right and stretch linearly in time.}
  \label{fig:horizontal_vs_vertical_setup}
\end{figure}

While we have discussed only cold plasma distributions so far, our method allows modeling of warm plasmas as well and we have have multiple ways of describing such warm plasmas. We will use the example problem of a Landau-damped electrostatic wave in a warm plasma to study the behavior of our method. In particular, we will discuss two extreme cases of setting up the sheet connectivity for modeling the Maxwellian velocity distribution. First, we will consider the perhaps more natural case with $M$ ``horizontal'' sheets of respectively constant but differing velocities set up parallel to the position axis. These sheets are evenly spaced in velocity space and the tracers within the sheets are evenly spaced in position space. We decrease the mass (and charge) of the particles in sheets with higher velocity to model the Maxwellian density in velocity space. In the second case, we consider $M$ ``vertical sheets'' starting out parallel to the velocity axis connecting particles with initially equal positions in position space across velocity space. Within the sheets, the particles' mass (and charge) decreases with velocity to model the Maxwellian distribution. We then perturb the warm distribution with a global velocity mode in order to study Landau
damping in this situation. Figure \ref{fig:horizontal_vs_vertical_setup} shows the phase space geometry and connectivity of the two approaches.

We use the ``cold start'' technique of loading $M$ cold sheets of particles in the PIC comparison as well. This is a standard approach \cite{schumer1998vlasov}  and stands in contrast to the ``warm start'', where the particles are distributed more randomly in velocity space. The trade-off is between an additional randomization of the particles, which fills in the gaps in velocity space better but incurs additional numerical noise --- which might overwhelm the electrostatic perturbation to be studied --- on the one hand, and the progression of the multi-beam instability on the other hand. For all methods, we expect our solutions to converge to the continuum solution as an increasing number of streams in velocity space are used \cite{dawson1960plasma}. 



\subsubsection{Convergence Study}

In order to quantify the error of our simulation\footnote{We perturb a warm plasma by a small velocity perturbation at the fundamental mode of the simulation region. The size of the region is chosen such that the thermal velocity is related to the phase velocity of the resulting wave as $v_{th}/v_{ph} = v_{th}/(L/T_p) = 0.6$. The perturbation is chosen with an amplitude $v_1 = 0.026 v_{th} \ll v_{th}$. The time step is the same for all runs as $dt = dx/v_{th}$, with $dx$ being the cell size of the gold run. We use $M = 20$ streams to sample velocity space uniformly between $\pm 4 v_{th}$.}, we perform one ``gold run''\footnote{$n_{ppc} = 200$, $n_{cells} = 1200$, $n_{streams} = 20$} and measure the potential energy as a function of time $E_{\rm{pot}}(t)$ over one plasma period. This gold run has high values for all relevant parameters determining accuracy, i.e. the number of particles per cell $n_{ppc}$ and the number of simulation cells $n_{cells}$. We then compare other solutions for $E_{\rm{pot}}(t)$ from other runs to this gold run by integrating the $L_1$ error over $t$. This allows us to study the error convergence rate. The gold run is performed with the standard PIC scheme. We perform this error convergence study for both piecewise constant segments and piecewise linear segments, as discussed in sections \ref{sec:segments} and \ref{sec:piecewise_linear}, respectively. The results are summarized in figure \ref{fig:ld_e_pot_error}.

As expected, the error decreases with increasing particle number for both schemes, until it converges to a minimum error produced by other factors, such as the the finite number of grid points. The figures show that the new scheme requires far less particles to converge to the minimum error than the old scheme does. The plot further suggests that for \emph{SIC} at low particle numbers, where the error due to the particle spacing is dominant, the error drops with a fixed power law, i.e. a fixed linear slope on the logarithmic plot.
Denoting the total number of particles per cell as $n_{ppc}$ and the error in the potential energy as $\epsilon_{E_{\rm{pot}}}$, we postulate
$$
\epsilon_{E_{\rm{pot}}} \propto n_{ppc}^{\gamma} \; ,
$$
where $\gamma$ denotes the rate of convergence of the method.

Note that the convergence rates as can be read off from the slopes of the error functions are consistent with first order convergence ($\gamma = -1$) for the piecewise constant segments and second order convergence ($\gamma = -2$) for the piecewise linear segments. The actual slope in the potential energy error plot is twice this value. To explain this, take $\tilde{E}_{\rm{pot}}(t)$ to denote our estimate in the simulation for the potential energy as a function of time. Moreover, let $\tilde{E}(x,t)$ denote our simulation estimate of the electric field, which we write as the true electric field $E(x,t)$ plus an error term $\epsilon(x,t)$. We then find that the first order error term integrates to zero and we are left with a term of twice that order:
\begin{eqnarray*}
\tilde{E}_{\rm{pot}}(t) &=& \int_{0}^{L} \tilde{U}(x,t) dx \propto \int_{0}^{L} (\tilde{E}(x,t))^2 dx \\
&=& \int_{0}^{L} (E(x,t) + h^d \epsilon(x,t))^2 dx \\
&=& E_{\rm{pot}}(t) + h^d \int_{0}^{L} E(x,t) \epsilon(x,t) dx + h^{2 d} \int_{0}^{L} \epsilon(x,t)^2 dx \\
&=& E_{\rm{pot}}(t) + h^d \langle E(x,t) \epsilon(x,t)\rangle_x +  h^{2 d} \epsilon_E(t)\\
&\approx& E_{\rm{pot}}(t) +  h^{2 d} \epsilon_E(t)
\end{eqnarray*}
In the first line, we take $\tilde{U}(x,t)$ as the potential energy density. In the second line, we keep only the lowest order error term, which we assume to be proportional to some step size $h$ at order $d$\footnote{In this case $h$ denotes the inverse number of particles per cell}. In the second to last line, we denote $\langle...\rangle_x$ as the average over $x$. Since the electric field and the error in the electric field are uncorrelated, this term approaches zero. This shows that it was OK to include only the lowest order term in the second line --- the linear term involving any higher order error would have integrated to zero as well and left a term of twice that order.

We also computed the error in the \emph{charge density} as a function of the number of particles per cell, and the results are consistent with those for the potential energy.

Several results are important to point out. Once the number of particles is large enough, the error is dominated by the finite grid spacing. As we increase the number of grid cells, this error baseline drops. It drops by the same amount for both schemes, showing that the primary source of improvement in the SIC scheme comes from reducing the ``shot noise'' associated with too few PIC particles. The smaller amount of particles required for accurate SIC simulations is an especially useful feature, since the number of particles is usually the major limiting computational factor for large-scale simulations. The errors produced by the grid spacing are consistent with second order convergence in $n_{cells}$, for both PIC and SIC. 

For piecewise constant segments in figure \ref{fig:ld_e_pot_error}, the error is lower than in the PIC simulation for the same number of particles per cell. Equivalently, far fewer particles than with PIC are required for the same error. For the piecewise linear segments, the effect is even more pronounced: the convergence rate is increased, and even fewer particles are required for the same error. Moreover, the minimum error for a given cell number is reached sooner. Since the \emph{total} number of particles per cell is measured by $n_{ppc}$, and these particles are distributed over all $M = 20$ streams, figure \ref{fig:ld_e_pot_error} shows that around one particle per cell in a given stream is sufficient for error convergence.

Finally, our data also shows that the error reaches a baseline (for $n_{cells} \approx 750$) at which it is limited by neither $n_{cells}$ nor $n_{ppc}$. We find that the error is now in fact dominated by the number of streams $M$, i.e. the coarse discretization of the distribution function in \emph{velocity space}.

\begin{figure}[!h]
\noindent\makebox[\linewidth]{
  \subfloat[Nonlinear Damping]{
  \includegraphics[width=0.42\linewidth]{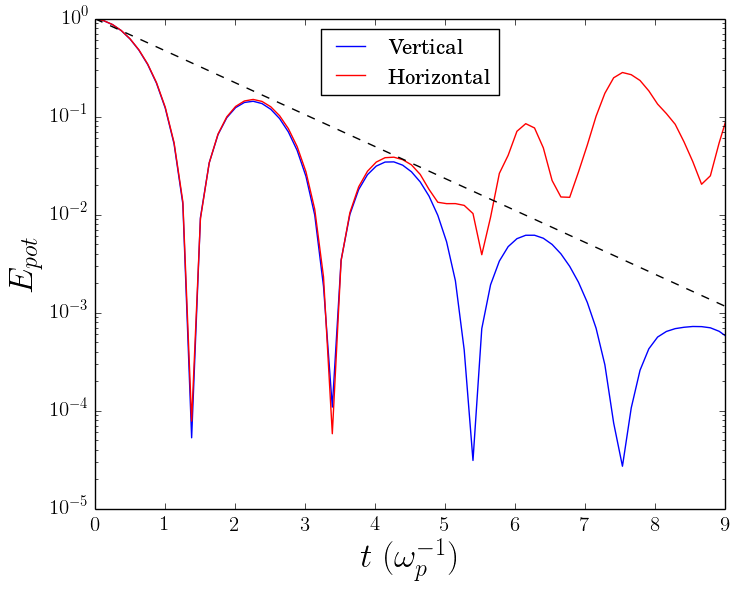}\label{fig:nonlinear_regime}
  }
  \subfloat[Linear Damping]{
    \includegraphics[width=0.5\linewidth]{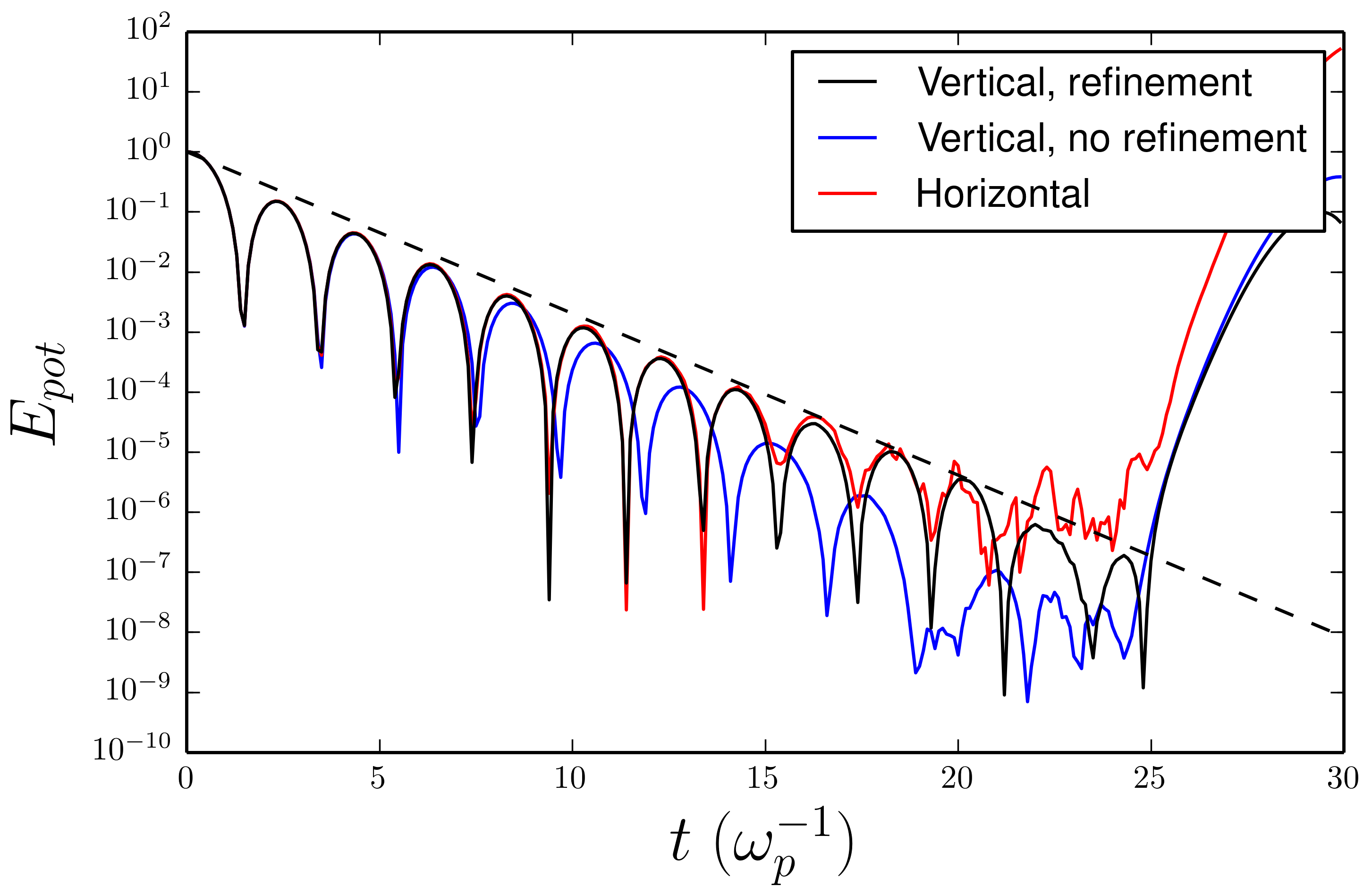}
    \label{fig:linear_regime}
    }
}

  \caption{We illustrate the tradeoff among different setups of the phase space geometry with the example of Landau damping with $M = 26$ streams and $160$ particles in every stream. We use the horizontal and vertical setups as shown in figure \ref{fig:horizontal_vs_vertical_setup}: the tracer particles have the same initial positions in both cases, but the connectivities are different (i.e. for the vertical space, we end up with $160$ vertical streams, with $26$ particles each). The horizontal streams tend to show noise once the multistream instabilities between the different streams develop. The vertical streams are not as affected by this. However, as the vertical streams stretch in phase space, the particle spacing becomes large and information is lost. Once the particle spacing becomes too large, noise develops as well. \protect\subref{fig:nonlinear_regime} A large velocity perturbation gives rise to nonlinear landau damping. We add an exponential trendline to guide the eye.  In this regime, the multistream instability develops rapidly and the horizontal streams become noisy. The vertical streams are not subject to this noise. \protect\subref{fig:linear_regime} A small velocity perturbation gives rise to linear landau damping. The horizontal streams remain without noise for a long time, while the vertical streams begin to show inaccuracies earlier (this is set by the time at which the initial particle spacing becomes comparable to the size of the box. However, as we add refinement, this problem is mitigated, and the vertical streams perform very well. For horizontal streams, the lack of refinement is never an issue, as the multistream instabilities would develop anyways. We also studied the above two cases using refined horizontal streams, and as expected the results were nearly identical to the unrefined case. This simple test problems illustrates that the phase space connectivity of the sheets is an algorithmic choice and can depend upon the problem in question.
}
  \label{fig:horizontal_vs_vertical}
\end{figure}

\subsubsection{Phase Space Geometry}
We performed an additional investigation to illustrate the flexibility and universality of dealing with connected sheets in phase space. So far, we had connected sheets along lines of constant velocity, so that the tracer spacing would stay small and constant for as long as possible. . However, this is not a requirement. To illustrate this, we perform the same simulation of Landau damping as before\footnote{We use the same parameters as in the section above, except that the initial sinusoidal perturbation is in position space. The amplitude is chosen to be small in the linear damping case at $4\times 10^{-4} dx$ and large in the nonlinear damping case $2 dx$. The simulation region has length $40dx$. The phase space is set up with $M = 26$ streams spaced evenly between $\pm 4 v_{th}$ and $160$ particles in each stream, i.e. $n_{ppc} = 160 \cdot 26 / 40$. The linear damping case is also performed using first-order weighting as described in section \ref{sec:first_order}. All other figures use the SIC deposit described in section \ref{sec:segments}.}, however, this time we compare horizontally connected sheets with (initially\footnote{Of course, vertical lines in phase space will deform into broader and flatter line structures as the tracers move at their respective velocities.}) \emph{vertically} connected sheets. A representation of the potential energy over time the runs is given in figure \ref{fig:horizontal_vs_vertical}. For horizontally connected sheets, the multistream instability \cite{dawson1960plasma} develops between the streams which produces noise in the solution. This instability is not as prominent when the sheets are vertically connected (they simply become broader and flatter lines). On the other hand, vertically connected sheets produce noise when they become too stretched in phase space and the particle spacing becomes comparable to the box size (most information is lost in this case). This problem can be avoided with refinement. Figure \ref{fig:horizontal_vs_vertical} shows that different test problems cause one method to perform better than the other. 

This toy example illustrates a key point. The connectivity in phase space of the sheets is an algorithmic choice. All simulations should in the limit approach the same solution. However, given finite computational resources, while it does not reflect any physical discretization, it is a choice that can affect the quality of the solution. There are many sensible ways of picking the connectivity, and in general the optimal connectivity will depend on the specific application.





\section{Discussion and Conclusions}\label{sec:discussion}
The SIC method represents an alternative discretization of phase space that offers merits over the traditional PIC method. In particular, for problems in which there is no turbulence or folding of sheets in phase space, SIC offers higher accuracy than PIC with significantly fewer tracer particles. Using first order information, SIC only requires the particle spacing to be sufficiently small to allow a reasonable piecewise linear approximation of the phase space density. Thus, a single particle for a few simulation regions is usually sufficient and at most one particle per simulation region is beneficial for error reduction, because the discretized simulation space cannot resolve lower-wavelength features either way. By contrast, PIC requires many tens of particles per simulation region. Since the simulation region spacing must itself be chosen to resolve the smallest scale of interest in the phase space distribution, this requirement results in orders of magnitude higher particle numbers. We have demonstrated this in several test problems, both for thermal and cold phase space distributions. Moreover, the computational effort per particle is nearly identical to PIC. Since particles are the main resource limitation for most large-scale simulations, a reduction in the required amount of particles is encouraging for the community.

Since the algorithms for PIC and SIC differ only in the charge deposit, it is worthwhile to consider a bit more closely the performance differences in this step. The cost for PIC is directly proportional to the number of particles, since each is deposited into a small and constant number of cells. For SIC, for $n_{ppc} < 1$ we can have one particle ``touching'' many cells, while a single particle will only touch a small number of cells if $n_{ppc} \gtrsim 1$ (since the average particle spacing is less than one cell). Thus, for $n_{ppc} > 1$ the cost is proportional to the number of particles as in PIC, while for $n_{ppc} < 1$, we have essentially a baseline cost equivalent to $n_{ppc} = 1$, because every cell needs to be touched, even when we have few particles. Despite this baseline cost, SIC can offer large savings over PIC for a given problem: as we have demonstrated --- for a fixed, small error tolerance --- SIC can be run effectively in the regime where $n_{ppc} \sim 1$, whereas PIC requires $n_{ppc} \gg 1$. Because in these regimes both have a computational cost $\sim n_{ppc}$ with a similar cost per particle, SIC offers significant savings. This is true in one dimension and increasingly so in higher dimensions. For nonlinear problems in which refinement for SIC becomes necessary, the cost of course increases over time compared to PIC. On the other hand, unlike PIC, SIC is able to resolve fine features in velocity space as long as this refinement is continued.

The key difference between SIC and PIC is that the functional approximation is not a sum of shape functions, but a set of tracers with an associated interpolation that approximates the functional form.
When folding and turbulent motion occur in phase space, the spacing between particles in the phase space sheets becomes unfavorable for an accurate deposit. In particular, as tracers move very far apart from each other, information about the distribution function between the tracers is lost. This can be corrected for with a refinement scheme. It is reasonable to make the refinement criterion based on the distance between tracers. We demonstrate the efficacy of refinement in problems with significant phase space folding and stretching. While refinement increases the computational cost of SIC simulations, the method in turn is able to capture the increasing complexity of the phase space distribution function over time as long as refinement is applied. This stands in contrast to other methods like PIC where this information is lost.

While our general method seems natural and intuitive in one dimension, it is also efficiently extensible to higher dimensions. In general, in $n$ spatial dimensions we are tracking connected phase space sheets that divide position space into $n$-simplices (for example, in $3$ spatial dimensions, we would have a division of position space into tetrahedra).

The phase space sheets in our method are functionally similar to the phase space contours in the so-called ``waterbag'' method \cite{colombi2014vlasov}. However, in $n$ spatial dimensions, the phase space sheets only define several manifolds of connected $n$-dimensional simplices, whereas the waterbag method requires manifolds of $2n$-dimensional polyhedra. Thus, our method --- with its reduced curse of dimensionality --- is easier extended to higher dimensions.

Encouragingly, it is possible to efficiently find the exact intersection of arbitrary polyhedra with cubic meshes \cite{franklin1993volumes}. This of course enables the exact SIC charge deposit step in higher dimensions. It is also possible to deposit higher dimensional polyhedra with linear fields onto cubic meshes \cite{powell2014intersection}. The gradient for the linear fields can be found either by fitting an $n + 1$ dimensional hyperplane to neighboring points, or by a generalization of finite differences \cite{correa2011comparison}. It is thus viable and realistic to efficiently extend SIC (including piecewise \emph{linear} segments) to higher dimensions. It is reasonable to assume that the sparsity of the representation of the distribution function will be even more beneficial for resource savings in these higher-dimensional spaces.

In general, SIC can be interpreted as removing some redundant information in the PIC description of phase space densities. The $2n$ dimensional functional form of the distribution function is approximated using a set of smooth $n$ dimensional manifolds. These smooth manifolds in return are approximated using a sufficient but not redundant number of tracer particles. This stands in contrast to PIC where the entire distribution function is Monte-Carlo sampled. From this perspective it is clear why the removal of redundant information results in higher efficiency encoding of the information contained in the phase space distribution function. However, as the true evolution of the distribution function surpasses the complexity that can be adequately captured by the set of available manifolds and their respective tracer particles, refinement is necessary in order to retain accuracy. Unrefined SIC should be viewed as an attempt of more efficiently encoding high-dimensional information.

While it of course does not represent a universal reduction in the complexity of representing the Vlasov-Poisson system numerically, SIC offers merits in terms of both physical accuracy and computational cost across a range of practical problems. 

\section*{Acknowledgements}
We thank Frederico Fiuza for useful discussions.  S.T. was supported by the Department of Defense (DoD) through the National Defense Science \& Engineering Graduate Fellowship (NDSEG) Program.

\bibliography{MyReferences} 

\end{document}